\documentclass{emulateapj}

\def\ga{\mathrel{\hbox{\rlap{\hbox{\lower4pt\hbox{$\sim$}}}\hbox{$>$}}}}
\def\la{\mathrel{\hbox{\rlap{\hbox{\lower4pt\hbox{$\sim$}}}\hbox{$<$}}}}

\usepackage{verbatim}
\usepackage{amsmath}
\usepackage{multirow}
\usepackage{relsize}
\usepackage{natbib}
\usepackage{color}
\usepackage{placeins}
\usepackage[utf8]{inputenc}
\usepackage[english]{babel}
\usepackage{appendix}
\usepackage[normalem]{ulem}

\usepackage{amssymb}
\usepackage{float}

\usepackage{scalerel,tikz}
\usetikzlibrary{svg.path}
\definecolor{orcidlogocol}{HTML}{A6CE39}
\tikzset{orcidlogo/.pic={
 \fill[orcidlogocol] svg{M256,128c0,70.7-57.3,128-128,128C57.3,256,0,198.7,0,128C0,57.3,57.3,0,128,0C198.7,0,256,57.3,256,128z};
 \fill[white] svg{M86.3,186.2H70.9V79.1h15.4v48.4V186.2z}
 svg{M108.9,79.1h41.6c39.6,0,57,28.3,57,53.6c0,27.5-21.5,53.6-56.8,53.6h-41.8V79.1z M124.3,172.4h24.5c34.9,0,42.9-26.5,42.9-39.7c0-21.5-13.7-39.7-43.7-39.7h-23.7V172.4z}
 svg{M88.7,56.8c0,5.5-4.5,10.1-10.1,10.1c-5.6,0-10.1-4.6-10.1-10.1c0-5.6,4.5-10.1,10.1-10.1C84.2,46.7,88.7,51.3,88.7,56.8z};
}}
\newcommand\orcidicon[1]{\href{https://orcid.org/#1}{\mbox{\scalerel*{
\begin{tikzpicture}[yscale=-1,transform shape]
\pic{orcidlogo};
\end{tikzpicture}
}{|}}}}

\shorttitle{Magnetic Field Orientation in GMCs}

\shortauthors{Barreto-Mota et al.}
\begin{document}
\title{\large{M\lowercase{agnetic} f\lowercase{ield} o\lowercase{rientation} \lowercase{in} s\lowercase{elf}-g\lowercase{ravitating} t\lowercase{urbulent} m\lowercase{olecular} c\lowercase{louds}}}

\vspace{2.0cm}

\author{Barreto-Mota, L. \altaffilmark{1}~\orcidicon{ }
\altaffiltext{1}{Instituto de Astronomia, Geofísica e Ciências Atmosféricas da USP, }}

\author{de Gouveia Dal Pino, E. M. \altaffilmark{1}~\orcidicon{ }
\altaffiltext{}{}}

\author{Burkhart, B.\altaffilmark{2,3}~\orcidicon{0000-0001-5817-5944} }
\altaffiltext{2}{Department of Physics and Astronomy, Rutgers University,  136 Frelinghuysen Rd, Piscataway, NJ 08854, USA}
\altaffiltext{3}{Center for Computational Astrophysics, Flatiron Institute, 162 Fifth Avenue, New York, NY 10010, USA}

\author{Melioli, C. \altaffilmark{4}~\orcidicon{ }
\altaffiltext{4}{Dipartimento Astronomia di Bologna}}

\author{Santos-Lima, R. \altaffilmark{1}~\orcidicon{ }
\altaffiltext{}{}}

\author{Kadowaki, L. H. S. \altaffilmark{1}~\orcidicon{ }
\altaffiltext{}{}}

\begin{abstract}

Stars form inside molecular cloud filaments from the competition of gravitational forces with turbulence and magnetic fields. The exact orientation of these filaments with the magnetic fields depends on the strength of these fields, the gravitational potential, and the line-of-sight (LOS) relative to the mean field. To disentangle these effects we employ three-dimensional magnetohydrodynamical numerical simulations that explore a wide range of initial turbulent and magnetic states, i.e., sub-Alfvénic to super-Alfvénic turbulence, with and without gravity. We use  histogram of relative orientation (HRO) and the associated projected Rayleigh statistics (PRS) to study the orientation of density and, in order to compare with observations, the  integrated density relative to the magnetic field. We find that in sub-Alfvénic systems the initial coherence of the magnetic is maintained inside the cloud and filaments form perpendicular to the field. This trend is not observed in super-Alfvénic models, where the lines are dragged by gravity and turbulence and filaments are mainly aligned to the field. The PRS analysis of integrated maps shows that LOS effects are important only for sub-Alfvénic clouds. When the LOS is perpendicular to the initial field orientation most of the filaments are perpendicular to the projected magnetic field. The inclusion of gravity increases the number of dense structures perpendicular to the magnetic field, reflected as lower values of the PRS for denser regions, regardless of whether the model is sub- or super-Alfvénic. The comparison of our results with observed molecular clouds reveal that most are compatible with sub-Alfvénic models.

\end{abstract}


\section{Introduction}\label{sec:intro}

Molecular clouds (MCs) are the cradles of star formation in our Galaxy. These are generally cold regions, with average temperatures often ranging between $10$ and $50$K, that exhibit filamentary structures produced by the interplay of supersonic magnetohydrodynamic (MHD)  turbulence and self-gravity \citep{2007ARA&A..45..339B}. The development of new observational techniques and several theoretical advances have helped with the understanding of several aspects of MCs, but still several questions are left unanswered \citep{2009ApJ...693..250B,2015ApJ...808...48B,2019FrASS...6...15P,2020SSRv..216...68G}. 

The presence of supersonic turbulence (with sonic Mach number $\mathcal{M}_s = v/c_s$ $>1$, where $v$ is the turbulent velocity and $c_s$ the sound speed), initially inferred through line width observations \citep{1981MNRAS.194..809L,1999ApJ...525..318P}, is one of the most important physical ingredients in these environments and has been extensively studied both theoretically \citep{doi:10.1146/annurev.astro.45.051806.110602,Krumholz07a,2011ApJ...730...40P,2013ApJ...770..150H,Federrath15a,Mocz2017} and observationally \citep{2004ARA&A..42..275S,2007ARA&A..45..339B,2010A&A...518L.102A,2014prpl.conf...27A}. Turbulence is also reflected in the structure hierarchy of these clouds and consequently in the statistical properties of the gas, such as the probability distribution functions (PDFs) of the density and column density field and the power-spectrum of density and velocity  \citep[see,][]{1959flme.book.....L,2008ApJ...686..350L,2009ApJ...692..364F,2012ApJ...750...13C,2013ApJ...763...51F,doi:10.1146/annurev.astro.45.051806.110602,2012A&ARv..20...55H}. The motion of the fluid can  be also affected by the presence of magnetic fields since ionized gas is also present and thus need to be accounted for to fully explain the dynamics of MCs \citep{2000ApJ...540..332P,2011MNRAS.414.2511V,2012ApJ...760...33L,2015ApJ...805..118B}.

The exact ratio between thermal pressure and magnetic pressure (i.e., the plasma beta parameter) or the Alf\'enic Mach number (i.e., the ratio between the velocity of the turbulence and the Alfv\'en velocity, $\mathcal{M}_A = v/v_A$) inside MCs  are not well-known, but there are several techniques that provide some insight into the intensity and morphology of the magnetic field. For example, the magnetic field intensity along a line of sight (LOS) can be estimated using the Zeeman effect, while its morphology in the plane-of-sky (POS) is more commonly observed using the polarized emission and absorption of light from asymmetric dust grains \citep{1993prpl.conf..279H,2012ARA&A..50...29C}. These grains are assumed to have their smaller axis aligned to the magnetic field, with the direction of the polarization of the thermal emission being perpendicular to the magnetic field, and the polarization from any background light that passes through the cloud and is absorbed by these aligned dust grains being parallel to the magnetic field. Observations are usually made in the visible and near-infrared spectrum, with more recent techniques also probing the polarization from the submilimiter spectra \citep{2013ApJ...768..159H,2014ApJS..213...13H,2013ApJ...769L..15S,2014Natur.514..597S,2014ApJ...797...74D,2015JAI.....450005H}.  The nature of this alignment is still a matter for debate, but evidence points to the action of radiative torques \citep[RATS, see][for further details]{2008MNRAS.388..117H,2018ApJ...852..129H}. More recently, another method has been proposed with the idea of velocity anisotropies in radio position-position-velocity channel maps, which can estimate the direction of the plane of the sky component of the magnetic field and also give a lower limit on the intensity of the magnetic field \citep{2014ApJ...790..130B,2015ApJ...814...77E,2015psps.book...81L,2016MNRAS.461.1227K}. 

Recently, observations made by \textit{Planck Telescope} \citep{2016AA...586A138P}, \textit{Herschel} and BLASTpol \citep{2013A&A...550A..38P,2017A&A...603A..64S}, have revealed the relation between the filamentary structures and the magnetic field of these regions. They have found that denser structures usually appear perpendicular to the magnetic field. \cite{2013A&A...550A..38P}, in particular, have observed not only large filaments in Taurus MC perpendicular to the magnetic field as well as less dense striations of gas parallel to the magnetic field which seem to be flowing towards the filament. Similar behavior has been observed by \cite{2016AA...586A138P}; \citep[see also][]{2019A&A...629A..96S} in several other clouds, with less dense structures appearing mainly aligned to the projected magnetic field in the plane of the sky.  

Previous studies have investigated the alignment between magnetic fields and density structures using 3D magnetohydrodynamic (MHD) simulations. \cite{0004-637X-774-2-128} and \cite{2017ApJ...842L...9H}  analyzed the alignment between structures and the magnetic field using statistical tools like those we will employ in this work. They found that most of the dense filaments are nearly perpendicular to the magnetic fields. This relation is also present in the column density maps. \citet{2017A&A...607A...2S} have  also investigated  the same issue and concluded that   the observed change in relative orientation between column density structures and the projected magnetic field in the plane of sky,  from mostly parallel at low column densities to mostly perpendicular at the highest column densities, would be the result of  gravitational collapse and/or convergence of flows. \cite{2020arXiv200300017S}, on the other hand, have recently found that a transition between structures parallel to the magnetic field to structures perpendicular to it not always can be seen, depending on the line of sight (LOS).

At smaller scales (i.e., down to 1000 AU scales), \cite{2017ApJ...842L...9H} and \citet{Mocz2017} also studied the morphology of the magnetic field around collapsed cores using ALMA data and 3D MHD simulations. They found that only in very strongly magnetized systems there is preservation of the field direction from cloud to disk scales and an hourglass-shaped field at scales smaller than 1000 AU. \cite{2018MNRAS.480.2939G} has also studied the collapse inside filaments through 3D MHD simulations. They noted that the magnetic field around the filament is primarily perpendicular to the structure and the collapse along the filament would later bend the magnetic field lines creating U-shapes. 

In this work, we extend upon these studies investigating how the alignment of density structures with the magnetic field is related to the presence of gravity and the sonic and Alfvénic Mach numbers associated to the turbulence  and how projection effects may affect observations for different lines of sight. For this aim, we performed and analysed several 3D MHD simulations with initial conditions compatible with observed MCs, with and without the inclusion of self-gravity.

This paper is structured as follows, in Section \ref{sec:methodology} we discuss the simulated models, their initial conditions and the statistical methods used to analyze our results. In Section \ref{subsec:hro} we present the results obtained from our models. In Section \ref{sec:comparison_obs} we compare these results with observations published in the literature and other numerical studies. In Section \ref{sec:dicussion} we discuss and compare our results with precious works and, finally, in Section \ref{sec:conclusion} we present our conclusions.

\section{Simulation setup}\label{sec:methodology}

In order to study the evolution of molecular cloud environments and how the initial stages of star formation are affected by turbulence, magnetic fields and self-gravity, we consider three-dimensional MHD simulations of two families of models. In the first set, the simulations only have the presence of MHD turbulence, with no self-gravity. These are the numerical simulations of isothermal molecular clouds performed by \citet{2003MNRAS.345..325C} and \citet{2009ApJ...693..250B}. The second set of models is composed by isothermal models with self-gravity. We run these new MHD simulations using a modified version of the code developed by \citep[see][]{2007ApJ...658..423K,2009MNRAS.394..157L,2010ApJ...714..442S}. Both sets of simulations have the similar initial conditions, as we will describe below. 

Our simulations solve the ideal MHD equations in their conservative form:

\begin{eqnarray}\label{eqtn_mhd_eqtn_method}
    \label{eqtn_mhd_eqtn_method:1}\frac{\partial \rho}{\partial t} + &\nabla& \cdot (\rho \textbf{v}) = 0 \\
    \label{eqtn_mhd_eqtn_method:2}\frac{\partial}{\partial t}(\rho\boldsymbol{v}) + \nabla \cdot \Big [\rho \boldsymbol{vv} &+& \Big ( P + \frac{B^2}{8\pi} \Big )\boldsymbol{I} - \frac{\boldsymbol{BB}}{4\pi} \Big ] = \rho \boldsymbol{g} + \boldsymbol{F}\\
    \label{eqtn_mhd_eqtn_method:4}\frac{\partial \textbf{B}}{\partial t} + \nabla &\cdot& (\boldsymbol{vB}-\boldsymbol{Bv})= 0\;, \quad \nabla \cdot \boldsymbol{B}=0
\end{eqnarray}

\noindent where $\rho$, $\boldsymbol{v}$ and $\boldsymbol{B}$ are density, velocity and magnetic field, respectively, $P$ is the thermal pressure of the gas, $\boldsymbol{g}$ is gravity, 
\textit{\textbf{I}} is the identity dyadic tensor and \textit{\textbf{F}} is the source term for the turbulence driving. We consider a cold cloud with a temperature of $10$K, which is of the order of the estimated temperature from cold dust emission inside cold dark clouds.

An ideal isothermal equation of state implies that the pressure can be written as $P = \rho c_s^2$, where $c_s = \sqrt{k_b T/\overline{m}}$ is the isothermal sound speed of the gas,  $\bar m$ is the average mass of the gas given by $\bar m= \mu m_H$, with the mean atomic weight $\mu=2.3$. 

In Eq. \ref{eqtn_mhd_eqtn_method:2}, the RHS term $\rho \boldsymbol{g}=-\rho \nabla \Psi$ is the gravitational force due to self-gravity. This term is considered only in the second family of models of our study. The gravitational potential $\Psi$ obeys the Poisson equation:

\begin{eqnarray}\label{eqtn_poisson_eq}
    \nabla^2 \Psi = 4 \pi G \rho.
\end{eqnarray}

Finally, the source term \textit{$\boldsymbol{F}$} appearing on the RHS term of Eq. \ref{eqtn_mhd_eqtn_method:2} is responsible for the turbulent energy injection. Turbulence is driven solenoidally continuously at every time-step. The forcing is calculated in Fourier space around a characteristic wavelength that defines the injection scale ($l_{inj} = 1/4 L$ for models with self-gravity and $l_{inj} = 1/2 L$ for models without self-gravity, where L is the size of the domain). The turbulent energy cascades down to the (numerical) dissipation scale within one dynamical time, $L/c_s$. Self-gravity is turned on after one dynamical time, which is at least two turnover times for the models with lower Mach numbers.

The first set of models, which neglects self-gravity, has $\psi = 0$ in the equations above. These simulations were built using a third order accurate hybrid essentially non - oscillatory (ENO) scheme to solve the MHD equations \citep{2003MNRAS.345..325C,2009ApJ...693..250B}, while the set of models with self-gravity employed a total variation diminishing (TVD) method \citep{2009ApJ...700...63K,2009MNRAS.394..157L,2010ApJ...714..442S}. To solve the Poisson equation, a multigrid method was used \citep[see][for further details]{2015MNRAS.448..207D}. Both codes are based on the Godunov method and use a Runge-Kutta procedure \citep[e.g.][]{2000ApJ...530..508L,refId0} for time integration.

\subsection{Boundary and initial conditions}\label{subsec:initial_cond}

\begin{table*}[t]
\caption{Initial conditions for all simulated models, with and without self-gravity}
\label{table_initial_param}
\begin{center}
\scalebox{1.2}{%
\begin{tabular}{lcccccccc}
\hline \hline
Simulation         & $\mathcal{M}_s$ & $\mathcal{M}_A$ & $n_0 (cm^{-3})$ & $\beta_0  =\frac{P_{th}}{P_{mag}}$ & $t_{ff} (Myrs)$ & Resolution & Turbulence & Gravity \\ \hline \hline
Turb\_Ms2.0\_Ma0.7 & 2.0             & 0.7             & 117.65        & 0.302                              & --------          & 256        & yes        & no      \\
Turb\_Ms4.0\_Ma0.7 & 4.0             & 0.7             & 444.81        & 0.080                              & --------          & 256        & yes        & no      \\
Turb\_Ms7.0\_Ma0.7 & 7.0             & 0.7             & 1779.25       & 0.020                              & --------          & 256        & yes        & no      \\
Turb\_Ms2.0\_Ma2.0 & 2.0             & 2.0             & 117.65        & 2.469                              & --------          & 256        & yes        & no      \\
Turb\_Ms4.0\_Ma2.0 & 4.0             & 2.0             & 444.81        & 0.653                              & --------          & 256        & yes        & no      \\
Turb\_Ms7.0\_Ma2.0 & 7.0             & 2.0             & 1779.25       & 0.163                              & --------          & 256        & yes        & no      \\
Grav\_Ms1.8\_Ma0.6 & 1.8             & 0.6             & 117.65        & 0.302                              & 7.23              & 512        & yes        & yes     \\
Grav\_Ms4.0\_Ma0.6 & 4.0             & 0.6             & 444.81        & 0.080                              & 3.72              & 512        & yes        & yes     \\
Grav\_Ms7.0\_Ma0.6 & 7.0             & 0.6             & 1779.25       & 0.020                              & 1.86              & 512        & yes        & yes     \\
Grav\_Ms1.8\_Ma2.0 & 1.8             & 2.0             & 117.65        & 2.469                              & 7.23              & 512        & yes        & yes     \\
Grav\_Ms4.0\_Ma2.0 & 4.0             & 2.0             & 444.81        & 0.653                              & 3.72              & 512        & yes        & yes     \\
Grav\_Ms7.0\_Ma2.0 & 7.0             & 2.0             & 1779.25       & 0.163                              & 1.86              & 512        & yes        & yes     \\ \hline
\end{tabular}%

}
\end{center}
\end{table*}

We consider a cubic Cartesian domain with periodic boundaries. For the first family of models, the units of length are given by the size of the injection scale ($L_{inj}$). The rms velocity ($\delta V$) is kept close to unity so that the velocity can be seen in units of $\delta V$, and $\boldsymbol{B}/(4\pi\rho)^{1/2}$ is the Alfv\'en velocity in the same units. The unit of time is given by the turnover time of the largest eddy, $L_{inj}/\delta V$. Density is in units of the initial ambient density $\rho_0$. The remaining quantity units are all derived from these ones. 

The simulations that consider self-gravity were performed using resolutions of $256^3$ and $512^3$ cells in the three directions of a uniform grid. Our lower resolution simulations were mainly used for testing and calibration of the initial conditions of the models that consider self-gravity. With that said, previous works have used similar resolutions of $256^3$ to $512^3$ for studying molecular clouds with good statistical convergence of the results already in $256^3$ resolution, at least for non self-gravitating turbulent models \citep[e.g.][]{2007ApJ...666L..69K,2009ApJ...693..250B,2010ApJ...714..442S}. The statistical behavior of models with a resolution of $256^3$ show very similar results to their high resolution counterparts (see Appendix \ref{sec:apendixA}). 

To evaluate the relative importance of turbulence compared to gravity, we will define the virial parameter, $\alpha_{vir}$, following \citet{1992ApJ...395..140B}:

\begin{equation}\label{eqtn_alpha_vir}
    \alpha_{vir} \sim \frac{2E_k}{E_g} = \frac{5 v_0^2 c_s^2 R}{G M},
\end{equation}

\noindent where $E_k$ and $E_g$ are the kinetic and gravitational energies of the system, respectively, $v_0$ is the one-dimensional rms velocity and $M$ is the mass evaluated over a sphere of radius $R$.

In the second set of simulations that consider gravity, units of length are given by the size of the domain ($L$), and units of velocity are given in terms of the isothermal sound speed ($c_s$), which implies time unit $L/c_s$. The magnetic field is scaled such that $\boldsymbol{B}_{c.u.} = \boldsymbol{B}/(4\pi)^{1/2}$. Density is given in units of initial ambient density $\rho_0$. 
The initial setup is built to ensure the minimum condition for collapse, such that the virial parameter (defined in Eq. \ref{eqtn_alpha_vir}) is  $\alpha_{vir} \sim 0.5$.

Our simulations are characterized by three parameters, the sonic Mach number $\mathcal{M}_s$ (with $\mathcal{M}_s = v/c_s$ being sonic Mach number and $v$ being the characteristic velocity of the turbulence), the Alfvénic Mach number $\mathcal{M}_A$ and, in the case of simulations with gravity, the virial parameter $\alpha_{vir}$ (Eq. \ref{eqtn_alpha_vir}). These parameters are described in Table \ref{table_initial_param}. The table also gives the corresponding initial thermal to magnetic pressure ratio ($\beta_0$) for each model, the initial free-fall time ($t_{ff} = (3\pi/32 G \rho)^{1/2}$), calculated for the initial $\rho_0$, for the simulations that consider self-gravity. We will  use this time scale to compare the evolution of different self-gravitating models.


Models without self-gravity are identified by the prefix \textit{Turb} and models that consider self-gravity are identified by the prefix \textit{Grav}. These simulations are drawn from the Catalog for Astrophysical Turbulence Simulations (CATS); \citep{2020arXiv201011227B}. It is important to note that there is a small difference in the parameters considered for the purely turbulent, non-gravitating models and the ones that consider self-gravity. Some of the models without self-gravity have $\mathcal{M}_A \sim 0.7$ , while corresponding models with self-gravity have $\mathcal{M}_A \sim 0.6$. Also, some models without self-gravity have $\mathcal{M}_s\sim2.0$, while the corresponding models with self-gravity, have $\mathcal{M}_s\sim1.8$. However, the differences are so small that the comparison between them is not compromised.

\subsection{General characteristics of the simulated models} \label{sec:general_models}

In this subsection we present the general visual characteristics of the models used in this study.

\begin{figure*}[ht]
\begin{center}
 \includegraphics[width=2.05
 \columnwidth,angle=0]{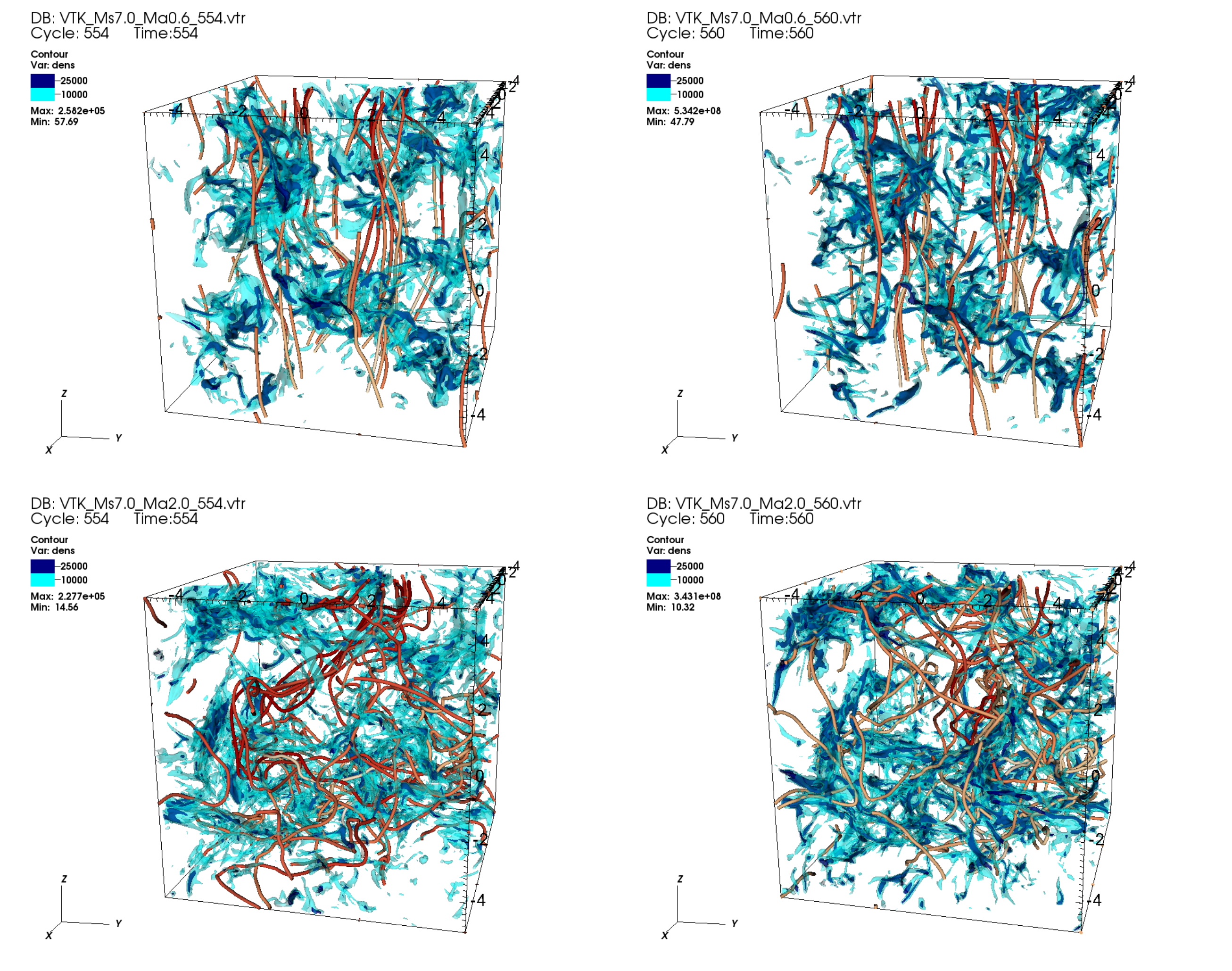}
    \caption{Comparison of the 3D distribution of density, represented by the iso-contours in light and dark blue, and magnetic fields, represented by the orange lines, for the initial ($t=0.0t_{ff}$; without self-gravity) and final snapshots ($t=0.3t_{ff}$; with gravity) for models \textit{Grav\_Ms7.0\_Ma0.6} (top) and \\textit{Grav\_Ms7.0\_Ma2.0} (bottom). The density contours are in units of $n/n_0$ (see Table \ref{table_initial_param}).}
    \label{fig_3D_Ms7.0_compare}
\end{center}
\end{figure*}

In Figure \ref{fig_3D_Ms7.0_compare} we show the three-dimensional views of the self-gravitating models before (left) and after (right) self-gravity is turned on. On the top of Figure \ref{fig_3D_Ms7.0_compare}, we show the supersonic, sub-Alfvénic model \textit{Grav\_Ms7.0\_Ma0.6} and in the bottom we show the supersonic super-Alfvénic model \textit{Grav\_Ms7.0\_Ma2.0}. When self-gravity is included (right-diagrams), fragmentation and filamentary structure formation is enhanced and the collapse of the densest regions (clumps) of these filaments leads to star formation.

{}

A closer inspection of the models of Figure \ref{fig_3D_Ms7.0_compare} shows that the distribution of the magnetic field lines is determined mainly by turbulence, and this effect is more pronounced in the case of super-Alfvénic turbulence. What determines if the lines become parallel or perpendicular to a given density structure, or if they are twisted inside the domain is whether the turbulence is sub-Alfvénic or super-Alfvénic. The twisting of the lines is more pronounced in the super-Alfvénic case due to larger turbulent energy relative to the magnetic energy. This also affects the filament structure, which appears more chaotically distributed with respect to magnetic field lines when the  magnetic field is weaker. In the case of the sub-Alfvénic model, we see that the lines are less distorted by turbulence and later on, when self-gravity becomes dominant, most of the filaments seem to be nearly perpendicular to the magnetic field lines (right top panel). In this case, collapse is not inhibited along the lines.

\begin{figure}[!h]
\begin{center}
\includegraphics[width=1.05 \columnwidth,angle=0]{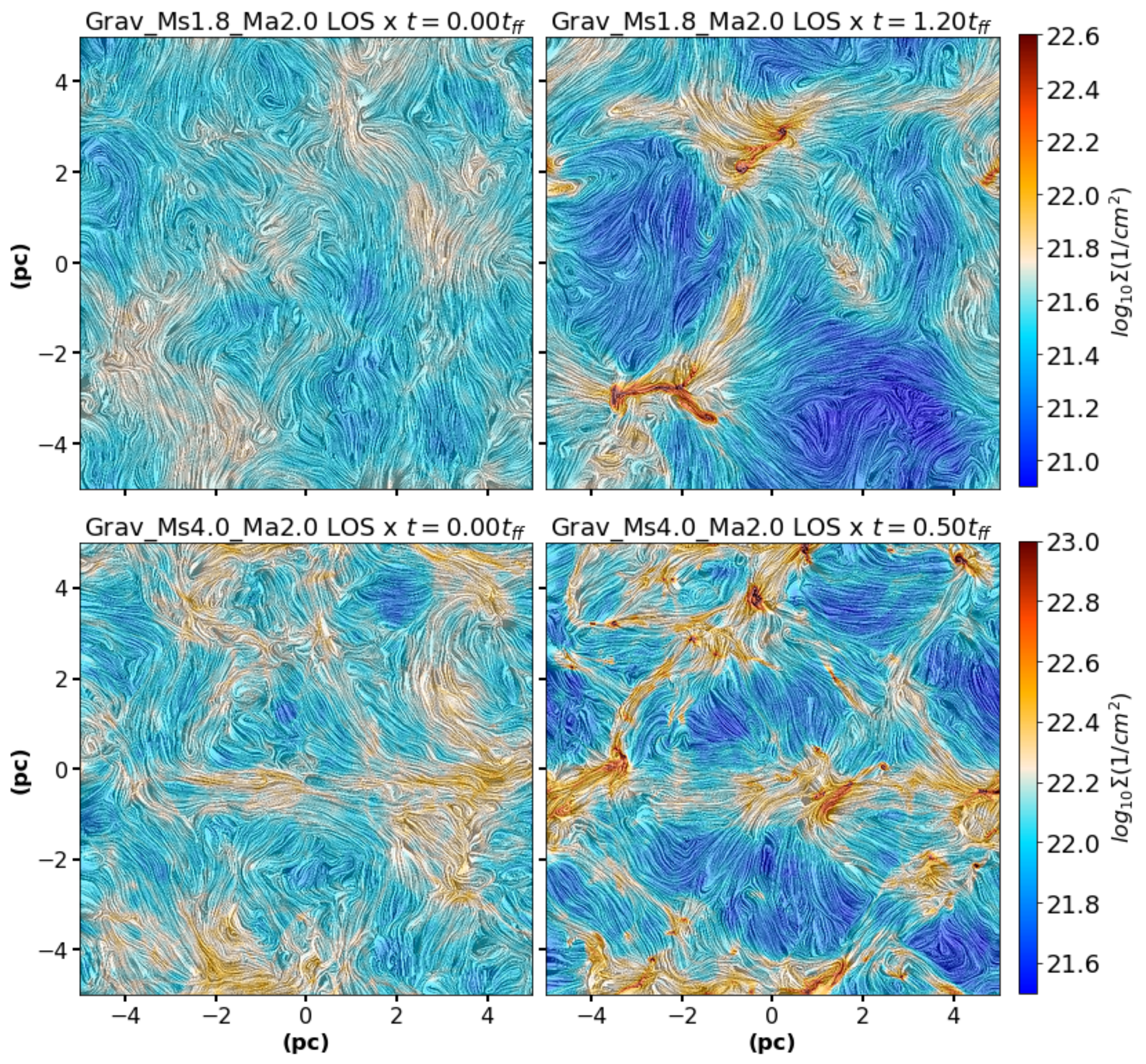}
\caption{
Column density maps with LIC method applied to $\boldsymbol{B}_\perp$ for  super-Alfvénic models with $\mathcal{M}_A= 2.0$ and  different sonic numbers ($\mathcal{M}_s= 1.8$, and $4.0$). In the left column we show the initial snapshot ($t=0.0 t_{ff}$), representative of models with fully developed turbulence without self-gravity. The right column presents the final snapshot of each model, at which  self-gravity has become important. In all panels we show the column density distribution integrated along X (the direction perpendicular to the initial field). The magnetic field is initially parallel to the Z axis. See Section \ref{sec:methodology} for further details.}
\label{fig:LIC_Ma2.0_evolution}
\end{center}
\end{figure}

\begin{figure*}[ht]
\begin{center}
\includegraphics[width=2.05 \columnwidth,angle=0]{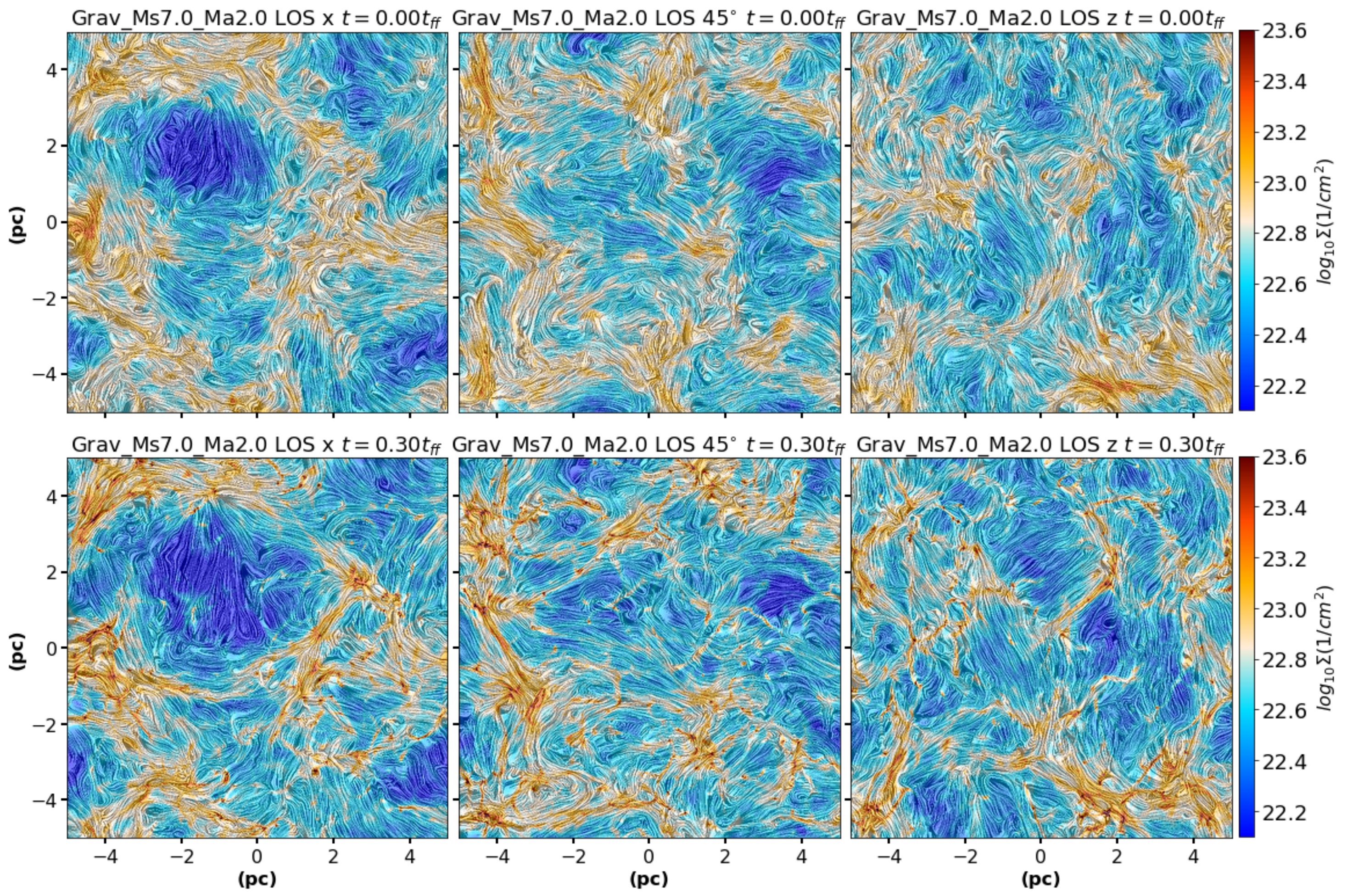}
\caption{Column density maps with LIC method applied to $\boldsymbol{B}_\perp$ for  super-Alfvénic models with $\mathcal{M}_A= 2.0$ and sonic Mach number $\mathcal{M}_s = 7.0$,  for three different LOS. Top row shows the column density map at the initial snapshot  ($t=0.0t_{ff}$) (with fully developed  turbulence  and no self-gravity), for each LOS. Bottom panel shows the final snapshot (when self-gravity becomes important) for the same LOS. From left to right the column density distribution is integrated along X (the direction perpendicular to the initial field), $45^\circ$ with regard to the initial field, and Z (parallel to the initial field) directions.}
\label{fig:LIC_Ms7.0_Ma2.0_evolution}
\end{center}
\end{figure*}

\begin{figure*}[ht]
\begin{center}
\includegraphics[width=1.98 \columnwidth,angle=0]{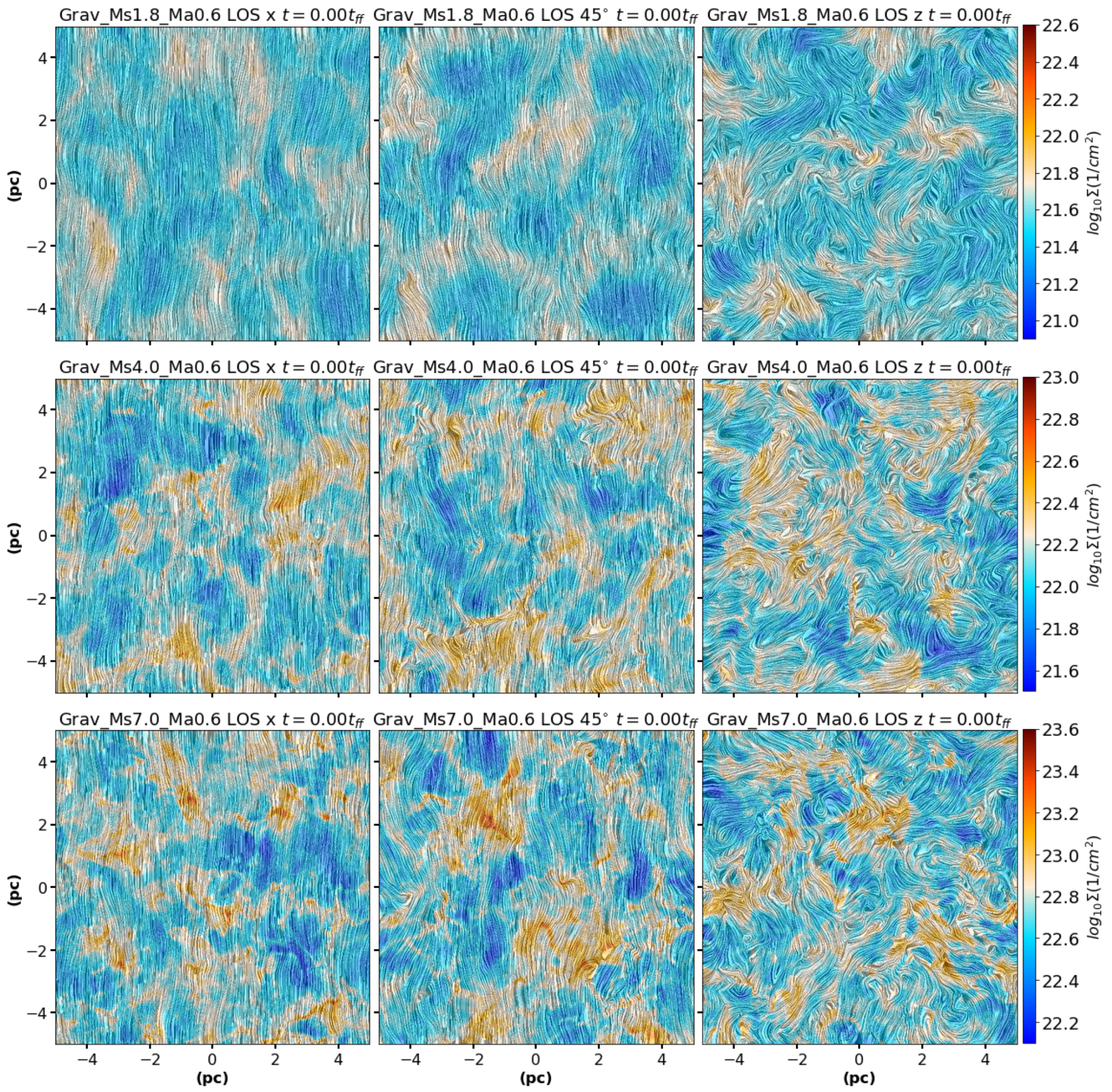}
\caption{Column density maps with LIC method applied to $\boldsymbol{B}_\perp$ for sub-Alfvénic models with  $\mathcal{M}_A= 0.6$ and different $\mathcal{M}_s$, at the initial snapshot (with fully developed turbulence and no self-gravity). From top to bottom panels  we show models with $\mathcal{M}_s= 1.8, 4.0$ and $7.0$, respectively. From the left to right panels, we show the LOS integration  along the X (the direction perpendicular to the initial field), $45^\circ$ with regard to the initial field and Z (parallel  to the initial field)  directions. }
\label{fig:LIC_Ma0.6_turb}
\end{center}
\end{figure*}

In Figures \ref{fig:LIC_Ma2.0_evolution} and \ref{fig:LIC_Ms7.0_Ma2.0_evolution} we present 2D  (column-density) maps of  super-Alfvénic models with  $\mathcal{M}_A = 2.0$,  in two different snapshots.
Figure \ref{fig:LIC_Ma2.0_evolution} compares models with different sonic Mach numbers, while Figure \ref{fig:LIC_Ms7.0_Ma2.0_evolution}, compares models with different LOS. The models in Figure \ref{fig:LIC_Ma2.0_evolution} consider $\mathcal{M}_s = 1.8, 4.0$ and $7.0$, as indicated on the top of each column density map. In the left column of this Figure, we show the initial snapshot $t=0.0t_{ff}$, corresponding to the time when the turbulence has completely evolved throughout the domain and has reached a steady state regime, before self-gravity is turned-on. In the right column, we show the final snapshot  for the same models, when self-gravity becomes dominant. We notice an increase in the formation of filamentary structures both with increasing $\mathcal{M}_s$ and with the introduction of self-gravity.

In Figure \ref{fig:LIC_Ms7.0_Ma2.0_evolution} all the models have initial $\mathcal{M}_s = 7.0$, and we show the  initial snapshot (with no self-gravity) in the top row and the final snapshot (when self-gravity has become dominant) in the bottom.

In  Figure \ref{fig:LIC_Ma2.0_evolution} all column density maps were integrated along a line of sight (LOS) perpendicular to the . Other LOS tests are presented in Figure \ref{fig:LIC_Ms7.0_Ma2.0_evolution}, but they all show very similar characteristics. None of the LOS have any distinctive characteristic, as one might expect for super-Alfvénic turbulence. The direction of the projected magnetic field in the plane of the sky ($\boldsymbol{B}_\perp$) is also shown and has been produced using a linear integral convolution method \citep[LIC,][]{cabral1993imaging}.

Figure \ref{fig:LIC_Ma0.6_turb} presents the column-density maps at initial snapshots (with fully developed turbulence and no self-gravity) of sub-Alfvénic models with $\mathcal{M}_A = 0.6$ and different sonic Mach numbers (from top to bottom, $\mathcal{M}_s = 1.8,4.0$, and $7.0$, respectively). From left to right, the Figure depicts   maps integrated along different LOS, namely, perpendicular, at an angle of $45^\circ$, and  parallel to the initial direction of $\boldsymbol B$, respectively.

\begin{figure*}[ht]
\begin{center}
\includegraphics[width=1.95 \columnwidth,angle=0]{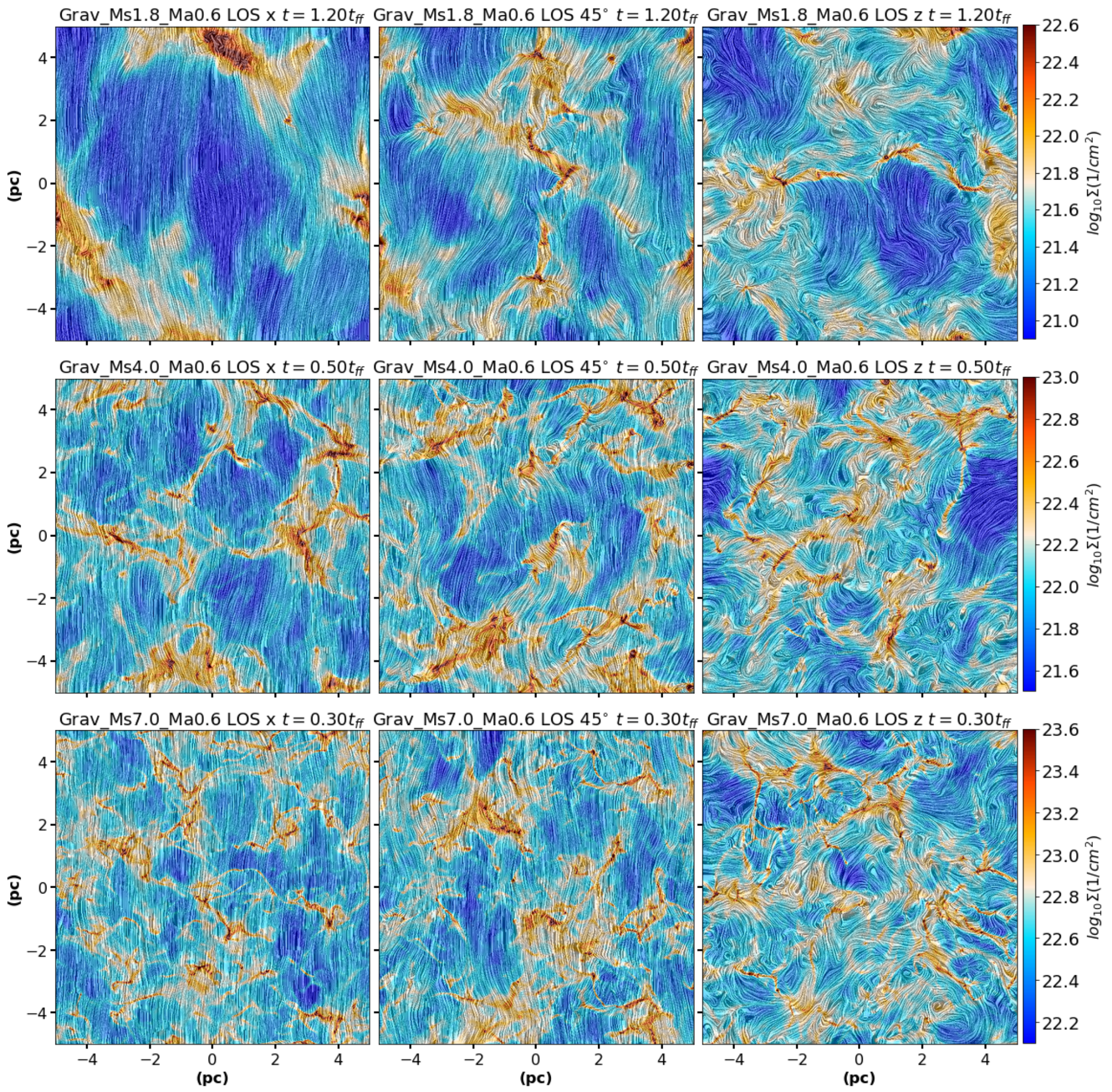}
\caption{Column density maps with LIC method applied to $\boldsymbol{B}_\perp$ for the same sub-Alfvénic models  presented in Figure \ref{fig:LIC_Ma0.6_turb}, but in an evolved time when self-gravity has become dominant. The snapshot time for each map is indicated on the top of each diagram. From top to bottom, we show $\mathcal{M}_s= 1.8, 4.0$ and $7.0$, respectively. From left to right, we show the LOS integration  along X (the direction perpendicular to the initial field), $45^\circ$ with regard to the initial field and Z (parallel to the initial field) directions. } 
\label{fig:LIC_Ma0.6_grav}
\end{center}
\end{figure*}

Figure \ref{fig:LIC_Ma0.6_grav}  shows the final snapshot (when self-gravity has become dominant) for the same sub-Alfvénic models as in Figure \ref{fig:LIC_Ma0.6_turb}, for comparison. Different from what we see in the super-Alfvénic models, the LOS here is important, changing the distribution of the $\boldsymbol B_\perp$, and influencing observed filaments.

The comparison between sub-Alfvénic and super-Alfvénic models shows that both depict dense filaments separated by diffuse interstellar gas, but these underdense structures when seen integrated along a LOS, in general, seem to be  larger (more coherent) in the super-Alfvénic models with no self-gravity. This effect is more pronounced in the LOS perpendicular to the initial $\boldsymbol B$ (see Figure \ref{fig:LIC_Ma2.0_evolution} left and top of Figure \ref{fig:LIC_Ms7.0_Ma2.0_evolution}). Supersonic turbulence leads to shocks and compression of gas and magnetic field lines, particularly in the early phase, before self-gravity sets in. In the super-Alfvénic case, these effects are more efficient in the building-up of large structures because the magnetic field strength is smaller than in the sub-Alfvénic models. In the latter, stronger magnetic pressure gradients offer larger resistance to the accumulation of the overdense structures by shock compression. When self-gravity becomes important, fragmentation and collapse will eventually dominate over the support provided by magnetic fields and turbulence in both cases, but the general imprints left earlier in the formation of the large scale filaments by turbulence and magnetic fields remain.

\subsection{General Statistics of the Simulations: Gravity versus Turbulence}

Supersonic turbulence in MCs can both enhance and inhibit over-dense regions that may eventually achieve sufficient conditions for collapse  
\citep{2006MNRAS.373..811M, 2009MNRAS.394..157L,2019ApJ...884L..35M}.
It is possible to study the interaction between turbulence and gravity through the use of statistical tools such as the one point and two point statistics, e.g., probability distribution functions (PDF) and the power spectrum \citep{2012ApJ...750...13C}, respectively.

In this subsection, we investigate the one and two point statistics of our two sets of simulations.  In this way we are able to benchmark them relative to theoretical expectations regarding  the interaction and transition of turbulent supported regions to self-gravitating collapsing regions.

\subsubsection{Density PDFs}
Due to interacting independent shock events, the isothermal density PDF in turbulent regions is  well represented by a lognormal distribution \citep{1994ApJ...423..681V,1997AIPC..393..101P,1998ApJ...504..835S,2018ApJ...863..118B}:

\begin{align}
    p(s) = \frac{1}{\sqrt{2\pi\sigma_s^2}}\exp\bigg(-\frac{(s-s_0)^2}{2\sigma_s^2}\bigg), \label{eqtn_lognorm_pdf}
\end{align}

\noindent where  $s \equiv \ln(\rho/\rho_0)$, $\sigma_s$ is the standard deviation of the lognormal, $s_0$ gives the value of $s$ for the mean density and can be related to the width as: $s_0= -\frac{1}{2}\sigma_s^2 $. The turbulent sonic Mach number is related to width of the lognormal as:

\begin{eqnarray}
    \sigma_s^2 &=& \ln[1+b^2 \mathcal{M}_s^2], \label{eqtn_sigma_lognorm}
\end{eqnarray}

\noindent where $b$ is a dimensionless turbulent forcing parameter \citep{2008ApJ...688L..79F} related to the solenoidal  and compressive modes of the turbulence.\footnote{For solenoidal turbulence driving, $\boldsymbol{\nabla}\cdot \delta v = 0$. For compressive turbulence driving, $\boldsymbol{\nabla}\times \delta v = 0$ \citep{2008ApJ...688L..79F}.} For purely solenoidal turbulence driving, we have $b=1/3$ while for purely compressive driving, $b = 1.0$.

Under the influence of self-gravity, the dense gas distribution of a turbulent cloud follows a power law $p_{PL}(s) \propto \exp(-\alpha s)$ for $s>s_t$, where $s_t$ is the transitional normalized density value between the diffuse gas given by the lognormal distribution and the power-law tail.  If the piecewise PDF is continuous and differentiable than \citep{2017ApJ...834L...1B}:

\begin{equation}
    s_t = \frac{1}{2}(2|\alpha| - 1) \sigma_s^2, \label{eqtn_rho_transition}
\end{equation}

\noindent where $\alpha$ is the power-law index.

The transition density between lognormal and powerlaw is related closely to the density at which gravity takes over the dynamics of the cloud, i.e., the so-called critical density for collapse \citep{2019ApJ...879..129B}. 
We consider that above a critical density $\rho_c$ all matter contributes to star formation. To estimate $\rho_c$ we 
consider the model presented in \citet{2011ApJ...730...40P}, which takes into account the contribution of the magnetic field for the determination of the critical density for star formation and can also be used for sub-Alfvénic turbulent conditions. The critical density is defined as:

\begin{equation}
    \frac{\rho_c}{\rho_0} = 0.067 \, \zeta^{-2}\alpha_{vir}\mathcal{M}_s^2 \frac{(1+0.925\,\beta^{-3/2})^{2/3}}{(1+\beta^{-1})^{2/3}},\label{eqtn_rho_crit_PN}
\end{equation}

\noindent where $\zeta \leq 1$ and $\zeta L_0$ is the turbulence integral scale ($L_0$ being the size of the system), and $\beta$ is the ratio between thermal pressure and magnetic pressure. In the limit $\beta \rightarrow \infty$, we recover the hydrodynamical case.

\begin{figure*}
\begin{center}
    \includegraphics[width=1.8 \columnwidth,angle=0]{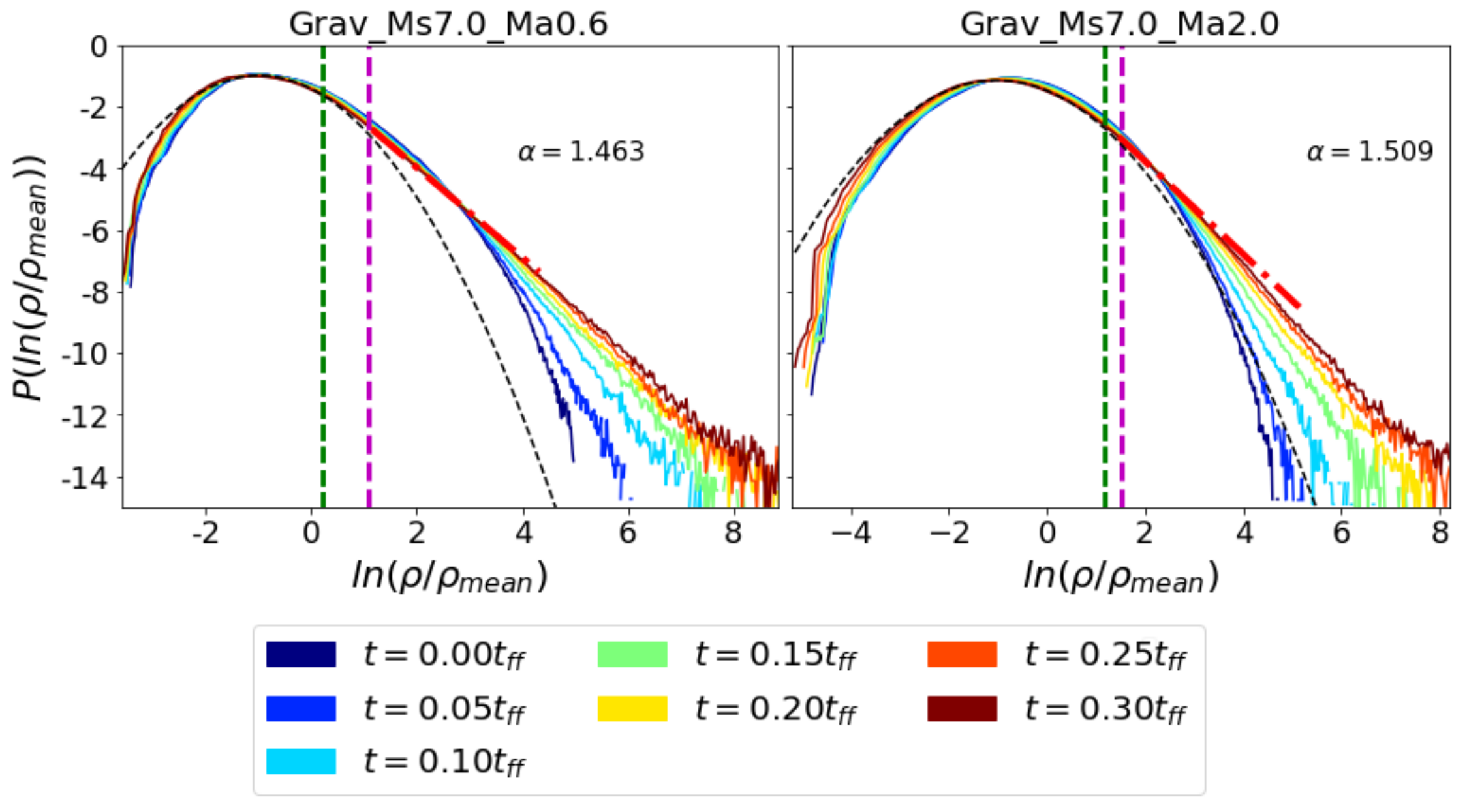}
    \caption{Time evolution of the density PDF for the super-Alfvénic  model\textit{Grav\_Ms7.0\_Ma0.6} (left) and the sub-Alfvénic model \textit{Grav\_Ms7.0\_Ma2.0} (right). The green vertical dashed line indicates the critical density for star formation (Eq. \ref{eqtn_rho_crit_PN}). The magenta vertical dashed line represents the transition density as defined in Eq. \ref{eqtn_rho_transition}. The black dashed line and the red dashed-dotted line are the fitted lognormal and power-law to $t=0.3t_{ff}$, respectively.}
    \label{fig_PDFs_Ms7.0_compare}
\end{center}
\end{figure*}

Figure \ref{fig_PDFs_Ms7.0_compare} compares the evolution of the PDF of the density (Eq. \ref{eqtn_lognorm_pdf}) for  two  models analyzed in the previous figures. The sub-Alfvénic model is presented in the left and the super-Alfvénic in the right panel. The blue dark curve represents $t=0.0t_{ff}$. Initially, the super-Alfvénic model shows a wider spread of density values when compared to the sub-Alfvénic model 
\citep[as in][]{2001ApJ...557..727V,2009ApJ...693..250B}. The black dotted line is the fitted lognormal PDF (Eq. \ref{eqtn_lognorm_pdf}) for $t=0.3t_{ff}$ and the red dashed-dotted line is the power-law fitted for the same time. The fitted index, and the region where the fit was considered, are indicated in the plot. The magenta vertical dashed line is the estimated $s_t$ (Eq. \ref{eqtn_rho_transition})  for the fitted power-law index and the green vertical dashed line is the critical density for the magnetized case (Eq. \ref{eqtn_rho_crit_PN}). We note that both coincide in the case of the super-Alfvénic model. As the material collapses, the power-law becomes shallower and eventually approaches $\rho_c$ \citep{2018ApJ...863..118B}.

The sub-Alfvénic model shows a small tail that deviates from the lognormal distribution, which is not present in the super-Alfvénic case. Since at this stage, gravity is not acting in the system, this deviation from the lognormal is most likely caused by the presence of a strong magnetic field \citep{2012ApJ...755L..19B}. At later times, the power-law reflects  the action of gravity, as discussed previously in the literature \citep{2011MNRAS.411...65B,2012ApJ...750...13C,2014ApJ...781...91G,2015ApJ...806..226M,2017ApJ...834L...1B,Mocz2017,2018ApJ...863..118B}.

Since our  code does not support a treatment of adaptive mesh refinement or sink particles, we consider that our results are valid until the power-law index reaches a value $\alpha\sim 1.5$. This has been chosen in accordance to previous studies that indicate power-law tails from observed clouds with an index up to this value \citep[see e.g., Table 1 from ][and references therein]{2018ApJ...863..118B}. In fact, the evolution of the cloud should result in a power-law with an index that converges to $-1$. However, effects due to the lifetime of the MC or to the LOS may yield steeper values for the observed power-law index \citep{2014ApJ...781...91G,2018MNRAS.477.5139G}.

\subsubsection{Power Spectrum}

The power spectrum can provide additional information about the development of turbulence present in the cloud. An incompressible fluid with fully developed hydrodynamical turbulence follows a Kolmogorov power spectrum ($P_{1D}(k)dk \propto k^{-5/3}dk$). However the slope of the spectrum may change in the presence of  magnetic fields, shocks and gravity. 

The collapse of structures due to the action of gravity produces very shallow slopes that may even become positive valued, similar to a delta-function  \citep{2013ApJ...763...51F,2015ApJ...808...48B}.

\begin{figure*}[!ht]
\begin{center}
\includegraphics[width=1.8 \columnwidth,angle=0,keepaspectratio]{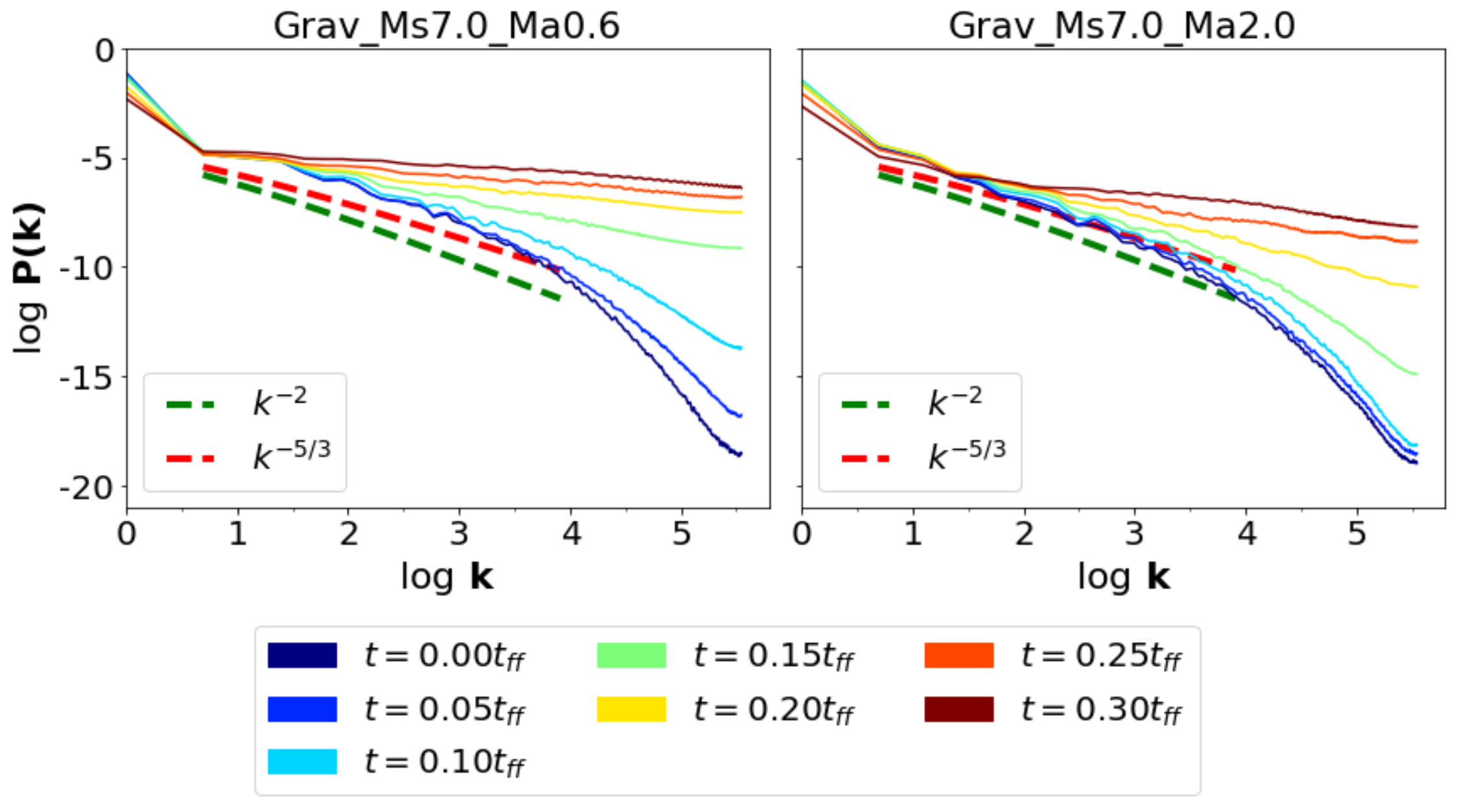}
\caption{Power spectrum of 3D density for models with $\mathcal{M}_s =7.0$ that do consider self-gravity. On the left we have $\mathcal{M}_A = 0.6$ and on the right we have $\mathcal{M}_A = 2.0$. The red dashed line represents the Kolmogorov power-law ($k^{-5/3}$) and the green dashed line represents the case for Burgers Turbulence ($k^{-2}$), for reference.}
\label{fig_pwrspec_3D_Ms7_compare}
\end{center}
\end{figure*}

Figure \ref{fig_pwrspec_3D_Ms7_compare} shows the evolution of the 1D power spectrum\footnote{The power spectrum was calculated along the x-axis (perpendicular to the initially homogeneous magnetic field). The power spectrum obtained along the other axes showed a very similar behavior.} of density evolution in time for the same two models analyzed in Figure \ref{fig_PDFs_Ms7.0_compare}. The dotted red line is a reference to the expected power-law from a Kolmogorov cascade ($P_{1D}(k)dk \propto k^{-5/3}dk$). Initially, in both simulations it is possible to identify an inertial region in the spectrum that roughly follows the same Kolmogorov slope.  As matter accumulates around  overdense regions, the power spectrum at higher wavenumbers $k$ (or smaller length scales) starts to flatten. This is a confirmation of the action of gravity at later times of the evolution of the cloud.


\section{Quantifying the relative orientation between density structures and the magnetic field}\label{subsec:hro}

In the previous sections we have demonstrated that our simulations show statistical and visual behavior compatible with previous works which  investigated the transition of supported turbulent gas to collapsing star forming gas. The primary aim of this work is to quantify the role that gravity and supersonic turbulence play in shaping the relationship between the magnetic field and collapsing dense structures. We focus on quantifying the importance of gravity vs. turbulence in the relative orientation between ISM density/column density  structures and the embedded magnetic field. 

The distribution or histogram of relative orientation (HRO) between the magnetic field and density is described in detail in \citet{0004-637X-774-2-128}.  The basics of the statistics are as follows: Gradients of density will be perpendicular to isodensity contours. Comparing the density gradient with the magnetic field orientation, it is possible to evaluate how the filamentary structures are aligned with the magnetic fields.

The relative orientation can be defined as:

\begin{equation}\label{eqtn_HRO_phi}
    \tan\phi = \frac{\textbf{B} \times \nabla\rho}{\textbf{B} \cdot \nabla\rho}.
\end{equation}

\noindent With $\phi$ as defined above, we can evaluate the histogram of $cos\phi$ within different density bins. A peak in this histogram around $cos\phi = 0$ means that  the field is perpendicular to the density gradient, in other words, the field is aligned with the structures. Similarly, $cos\phi = \pm 1$ means the density gradient is parallel to the field, hence the field is perpendicular to the structures. 

The HRO method can also be applied to the column density gradients, much in the same way, the only difference is the fact that the histogram is generally evaluated for $\phi+90^{\circ}$ in this case. In three dimensions, two random vectors have a higher probability of being perpendicular to each other  than being parallel, that is why we choose to use $\cos\phi$ in 3D. This is not the case in two dimensions and we can use simply the information from $\phi$. This is also what is commonly used in observations, e.g. Planck \citep{2016AA...586A138P,2017A&A...603A..64S}.

For observational data, the angles are calculated between the gradient of column density vector and the estimated $B_{\perp}$ from polarization data obtained from the Stokes parameters, i.e., a set of values that characterize the polarization of electromagnetic waves.

In order to compute these parameters for polarized radiation from the simulations, we will use the same assumptions as \citet{2008ApJ...679..537F}; \citep[see also][]{2015HiA....16..391P}. We assume that only thermal emission is emitted by grains that are perfectly aligned with the magnetic field. The dust abundance (which is not explicitly accounted for in our simulations) is considered to be proportional to the gas density, as well as the intensity of its emission. Finally, we consider that all grains emit at a single temperature. With these assumptions, for each cell of the computational domain we calculate:

\begin{equation}\label{eqtn_stokes_parameters}
    \begin{split}
     q = \rho \: \cos\, 2\psi \: \sin^2  i,  \\
     u = \rho \: \sin\, 2\psi \: \sin^2  i,
    \end{split}
\end{equation}

\noindent where  $\rho$ is the local density, $\psi$ is the local angle of alignment, determined by the projection of the local magnetic field onto the plane of the sky, and $i$ is the angle between the magnetic field and the line of sight (LOS). Integrating $q$, $u$ and $\rho$ along a chosen line of sight results on the Stokes parameters $Q$, $U$:

\begin{equation}\label{eqtn_stokes_parameters_integrated}
\begin{split}
    Q &= \int q \,dl, \\
    U &= \int u \,dl, \\
\end{split}
\end{equation}

\noindent and the column density:

\begin{equation}
    N_H = \int \rho \,dl.  \\
\end{equation}

This way, the intensity of the polarization vector $E$ and its direction will be calculated as:

\begin{equation}
\begin{split}
    \label{eqtn_B_pos}
    E &= \sqrt{Q^2+U^2}, \\
    \Phi_B &= 1/2 \; \arctan(U/Q).
\end{split}
\end{equation}

We can use Equation \ref{eqtn_HRO_phi}, where $\boldsymbol\nabla \rho$ can also be replaced by $\boldsymbol\nabla N_H$ and $\textbf{B}$ by $\textbf{B}_{\perp}$ if we use column density and polarization maps, to build a histogram to evaluate the angular distribution. 

To analyze the behavior of filaments inside molecular clouds we need to evaluate a wide range of density values that spans 2 to 3 orders of magnitude. To check how the alignment between density structures and the magnetic field occurs at different scales, we analyze the density information in several bins. To guarantee comparable statistics for each density bin, the density range is divided into bins with the same number of grid cells and then the HRO is calculated for each bin. 

In order to study the relative orientation between the (column) density gradient and the (projected) magnetic field, an additional statistical method can be used, namely, the projected Rayleigh statistics or PRS.
This can be calculated as \citep{2018MNRAS.474.1018J}:

\begin{equation}\label{eqtn_PRS}
    Z_x = \frac{\Sigma^n_i \cos\theta_i}{\sqrt{n/2}},
\end{equation}

\noindent where $\theta_i \in [-\pi,\pi]$ is the set of angles between the two vector quantities that we want to characterize and $n$ is the total number of angles in our set. Positive values of $Z_x$ are indicative of strong parallel alignment between the two vectors, while negative values indicate a strong perpendicular alignment between them. 

\citet{2018MNRAS.474.1018J} argue that, in the limit of $n \rightarrow \infty$, the PRS approximates the standard normal distribution. Therefore, for a general distribution of angles, the variance of $Z_x$ can be estimated as 

\begin{equation}\label{eqtn_error_PRS}
    \sigma^2_{Z_x} = \frac{ 2 \; \Sigma^n_i (\cos\theta_i)^2 - (Z_x)^2 }{ n }. 
\end{equation}

The error of each measurement of $Z_x$ will be given by the equation above.

Equation \ref{eqtn_PRS} cannot be applied to a 3D distribution, but  following Chapter 10 of \cite{mardia1999tests}, we can perform a  similar test to calculate  $Z_{3D}$ as:

\begin{equation}
    \begin{split}
        Z_{3D} = 2\,k\,\Big(\sqrt{\Big(\Sigma^n_i{\sin\theta_i}\Big)^2 + \Big(\Sigma^n_i{\cos\theta_i}\Big)^2} - \Sigma^n_i{\cos\theta_i}\Big),\label{eqtn_PRS_3D}
    \end{split}
\end{equation}
where $k$ is called the concentration parameter. As the name suggests, it is a weight parameter that describes the concentration of vectors around the mean direction. For our calculations, we will consider $k=1$. As we will see later, the absolute value of $Z_{3D}$ is not as important as the variation of $Z_{3D}$ for different density contours. 


When analyzing the density distribution of our 3D simulated cube, we will calculate the PRS values for the angles between the gradient of density ($\boldsymbol{\nabla} \rho$) and the magnetic field ($\boldsymbol{B}$). Large values indicate that $\boldsymbol{B}$ is perpendicular to $\boldsymbol{\nabla} \rho$ while values closer to zero indicate that these two vectors are parallel. 

For the integrated density maps along a given line of sight (column density), what is being evaluated is the angle between the polarization pseudo-vector $E$ (instead of the magnetic field projected on the plane of the sky, $\boldsymbol{B_\perp}$) and the gradient of column density $\boldsymbol{\nabla} N_H$. In this case Eq.\ref{eqtn_PRS} will be referred as $Z_{2D}$. Given the isocontour of a column density map, since $\boldsymbol{\nabla} N_H$ is perpendicular to it and $\boldsymbol{E}$ is perpendicular to the projected magnetic field in the plane of the sky, $\boldsymbol{B}_\perp$, the  angle between them,  $\phi$, will be also between $\boldsymbol B_\perp$ and the direction of the density structure (iso-contour). With this in mind, if $\boldsymbol{E}$ is parallel to the gradient of column density $\boldsymbol{\nabla}N_H$, then $Z_{2D} > 0$, and if perpendicular, $Z_{2D} < 0$.



\subsection{Results of the PRS and HRO}\label{sec:prs_hro_results}

In this work, all the PRS calculations considered 20 bins, both for $Z_{3D}$ and $Z_{2D}$. Tests with additional bins did not add any relevant information. As we will see, the information provided by both criteria $Z_{3D}$ and $Z_{2D}$ must be seen as complementary to each other to provide a whole picture of the relative distributions of the structures and their magnetic fields.

\subsubsection{Models without self-gravity}\label{subsec:turb_sims}

\begin{figure}[!h]
\begin{center}
\includegraphics[width=1.05 \columnwidth,angle=0]{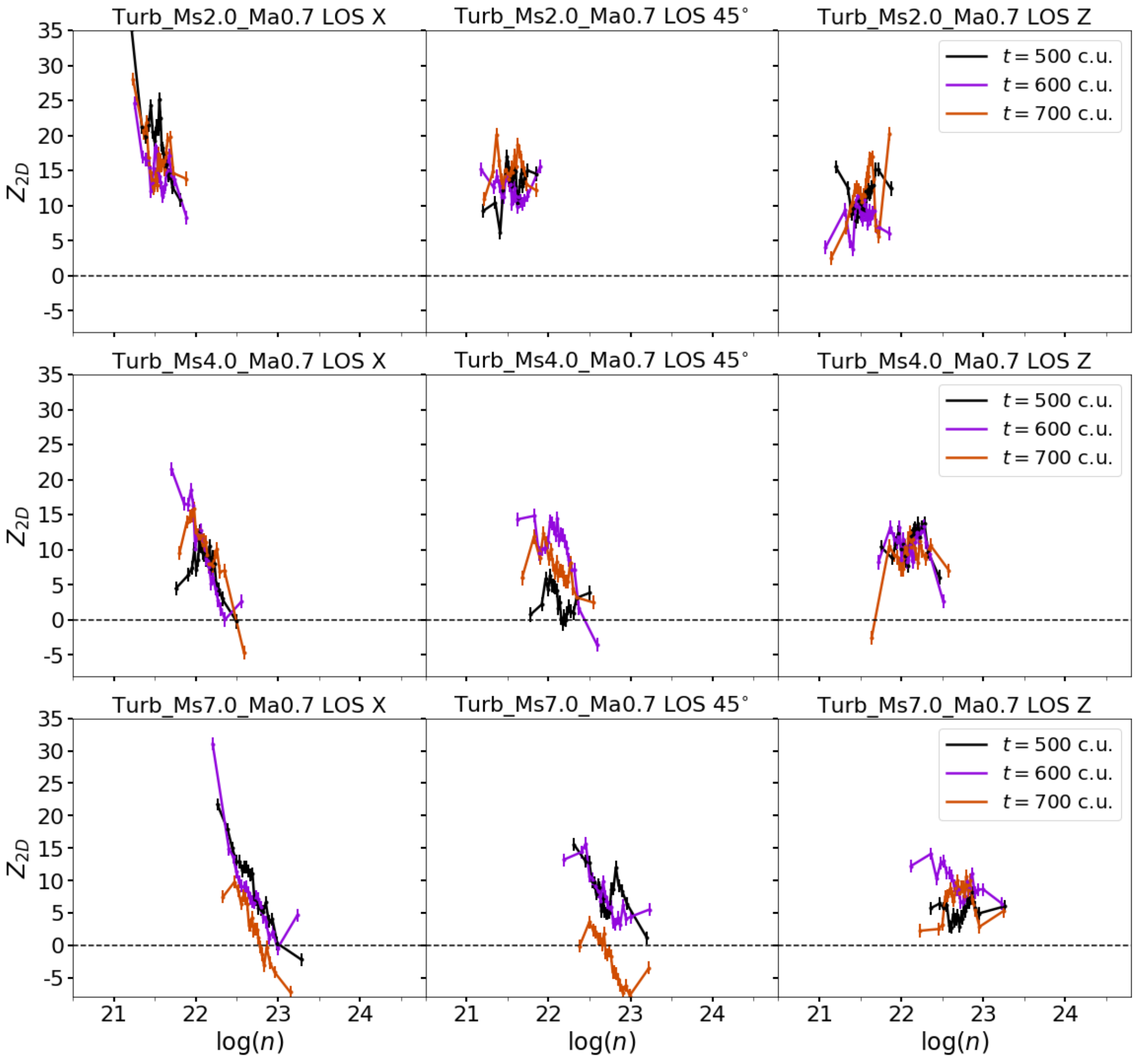}
\caption{PRS time evolution for all sub-Alfvénic models (with $\mathcal{M}_A=0.7$) without self-gravity. From left to right the PRS is applied to the LOS along X (the direction perpendicular to the initial field), $45^\circ$ with regard to the initial field and Z (parallel to the initial field) directions. From top to bottom initial $\mathcal{M}_s= 2.0, 4.0, 7.0$, respectively. }
\label{fig_2D_all_hros_godunov_Ma0.7}
\end{center}
\end{figure}

Figure \ref{fig_2D_all_hros_godunov_Ma0.7} shows the PRS analysis for all the sub-Alfvénic models and for the three different lines of sight. First, when the LOS is perpendicular to the initial uniform field, $Z_{2D}$ has a broader variation in time for higher densities. 

Positive values of $Z_{2D}$ indicate that $\boldsymbol{E}$, the direction of the polarization vector, is parallel to $\boldsymbol \nabla N_H$, which indicates that the projected magnetic field in the plane of sky ($\boldsymbol B_\perp$) is parallel to the iso-contours of $N_H$. When LOS is parallel to the initial magnetic field, the PRS analysis returns only positive values for most densities with very little variation. This happens because $\boldsymbol{B}_\perp$ in this case results from motions perpendicular to the main component of the field and this results in a random field distribution as seen in Figures \ref{fig:LIC_Ma0.6_turb} and \ref{fig:LIC_Ma0.6_grav}. This is the case for the maps on the right column diagrams from Figure \ref{fig:LIC_Ma0.6_grav}, which show that the coherence length is smaller in this LOS. Since we are projecting only the plane of the sky component of the field, we do not see the main component, only perturbations perpendicular to it.

In the other two LOS of Figure \ref{fig_2D_all_hros_godunov_Ma0.7}, as the sonic Mach number is increased (keeping the same $\mathcal{M}_A= 0.7$), we identify some negative values in the PRS for the largest column densities. This indicates that, when observed from these LOS, the densest structures tend to be \textbf{less} aligned with the local projected magnetic field to the plane of sky. This seems to be counter-intuitive to what one should expect, since this effect seems to be larger for larger turbulent motions (larger $\mathcal{M}_s$) relative to the magnetic field strength, where compression effects should be even stronger. However, looking at the column density maps of Figures \ref{fig:LIC_Ma2.0_evolution} through \ref{fig:LIC_Ma0.6_grav}, we note that the increase of turbulence (increase of $\mathcal{M}_s$) causes more fragmentation and the formation of more numerous smaller and denser structures. This effect is more pronounced for the LOS along \textit{X} (where the projected magnetic field to the sky has a larger component aligned to the original magnetic field) and less pronounced as we go to the LOS along \textit{Z} (where the projected magnetic field to the sky has a smaller component aligned to the original magnetic field). In other words, in these sub-Alfvénic models, only the densest and smallest structures that develop from increased fragmentation in the more supersonic (larger $\mathcal{M}_s$) at latter stages of evolution, tend to align with the intrinsic magnetic fields, and this effect is observable only for LOS perpendicular or with angles around $45^\circ$ to the original field.

\begin{figure}[!h]
\begin{center}
\includegraphics[width=1.05 \columnwidth,angle=0]{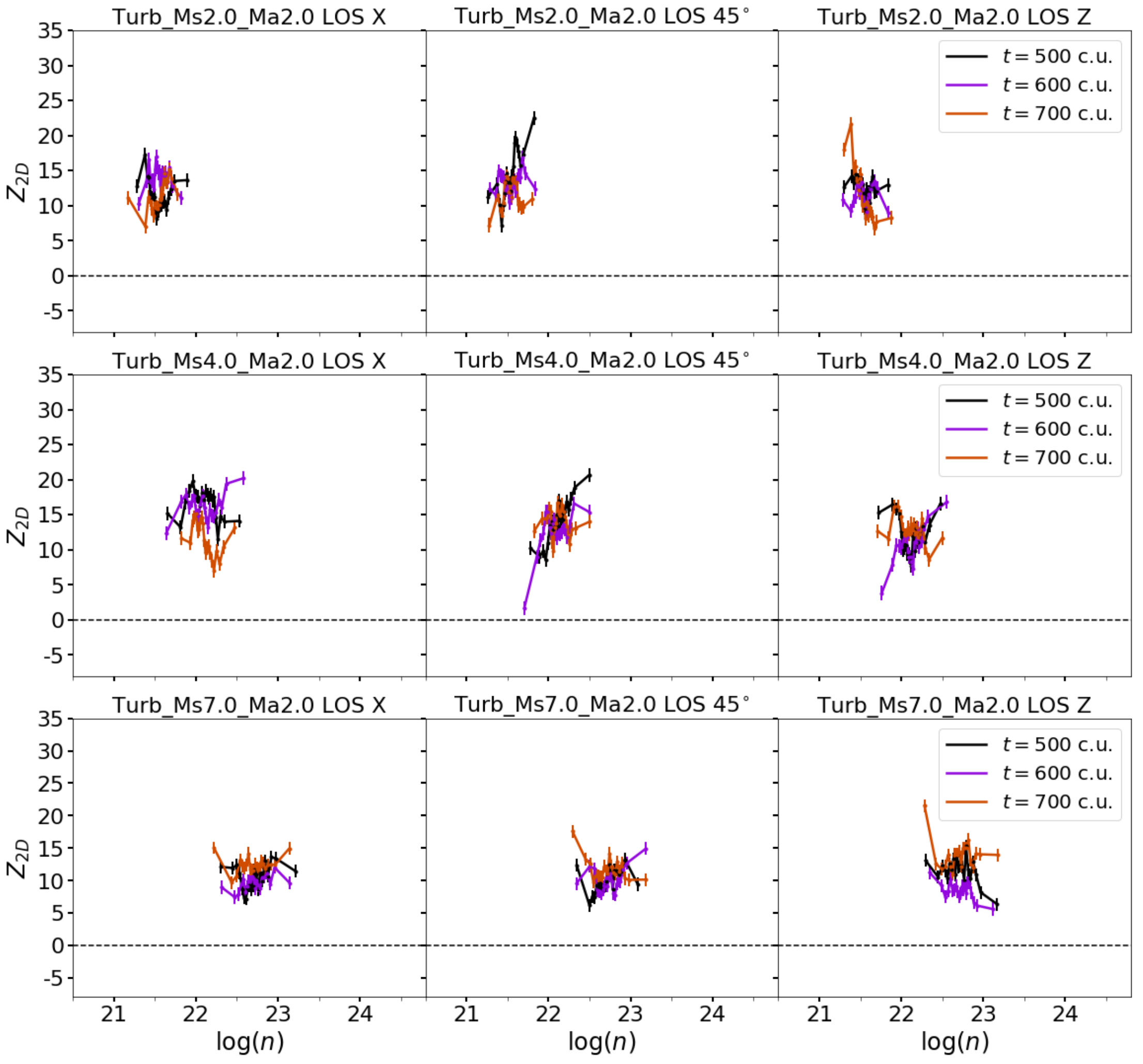}
\caption{PRS time evolution for all super-Alfvénic  models (with $\mathcal{M}_A=2.0$)  without self-gravity. From left to right the PRS is applied along LOS X (the direction perpendicular to the initial field), $45^\circ$ with regard to the initial field and Z (parallel to the initial field) directions. From top to bottom initial $\mathcal{M}_s= 2.0, 4.0$ and $7.0$, respectively} 
\label{fig_2D_all_hros_godunov_Ma2.0}
\end{center}
\end{figure}

Figure \ref{fig_2D_all_hros_godunov_Ma2.0} shows similar PRS analysis as in Figure \ref{fig_2D_all_hros_godunov_Ma0.7}, but for  the  super-Alfvénic models. When $\mathcal{M}_A = 2.0$ (therefore, decreasing the strength of the magnetic field relative to the turbulent motions), the PRS does show positive values and thus aligned structures to the projected magnetic field for all LOS (see Figure \ref{fig_2D_all_hros_godunov_Ma2.0}). In this case the magnetic field is not strong enough and, while  wandering due to the turbulence, it is also compressed by the supersonic flow, following the fragmented filaments. For most of the super-Alfvénic cases the behavior of the PRS is closer to what is seen in the maps integrated along \textit{Z}  in the sub-Alfvénic case. Therefore, in general, the magnetic field presents itself parallel to the filaments in these super-Alfvénic, supersonic flows.

\begin{figure}[!h]
\begin{center}
\includegraphics[width=1.05 \columnwidth,angle=0]{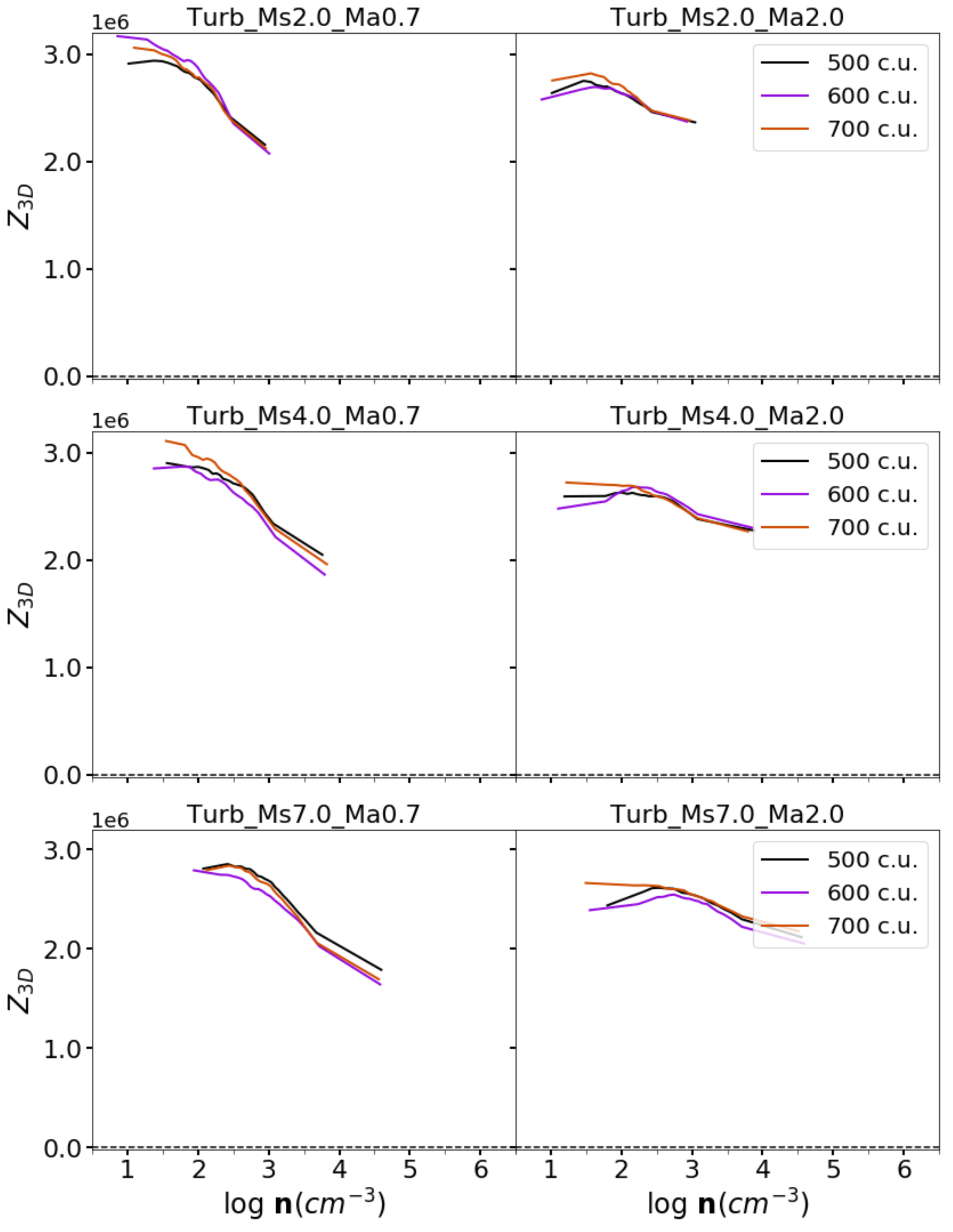}
\caption{$Z_{3D}$ analysis 
(Equation \ref{eqtn_PRS_3D})
applied to the 3D distribution of density and magnetic fields for all simulated models without self-gravity (see Table \ref{table_initial_param}) in three different snapshots indicated in the inset. On the left are presented the sub-Alfvénic (initial $\mathcal{M}_A=0.7$), and on the right super-Alfvénic (initial $\mathcal{M}_A=2.0$) models. From top to bottom initial $\mathcal{M}_s= 2.0, 4.0$ and $7.0$, respectively.}
\label{fig_3D_all_hros}
\end{center}
\end{figure}

Figure \ref{fig_3D_all_hros} depicts the $Z_{3D}$ analysis for the models without self-gravity. 
This analysis method indicates that in general the  density gradient is predominantly perpendicular to $\boldsymbol{B}$ (and hence, structures are actually mostly aligned with the intrinsic magnetic fields), since $Z_{3D}\gg0$ for all densities. This applies to both sets of simulations, sub-Alfvénic and super-Alfvénic. The sub-Afvénic models show 
smaller values of $Z_{3D}$ towards denser regions compared to their super-Alfvénic counterparts. These results indicate that the compression of the lines by the gradient of pressure is the dominant factor due to the supersonic turbulence. In the super-Alfvénic case, the compression of the magnetic field lines in the direction parallel to the density (pressure) gradient  and perpendicular to $\boldsymbol{B}$ is more effective, resulting in the alignment of the magnetic field with the  filaments.  In the sub-Alfvénic case, the more intense $\boldsymbol{B}$   offers greater resistance to compression.

This $Z_{3D}$ analysis seems to be, at first sight, a little in contradiction with the $Z_{2D}$ analysis performed before for the column density distributions, at least for the densest structures. However, one has to have in mind that  $Z_{2D}$ is subject to projection effects, while the $Z_{3D}$ analysis provides intrinsic values. Indeed, the fact that only positive values of $Z_{3D}\gg0$ appear does not mean that there are no regions where $\boldsymbol{\nabla}\rho$ is parallel to $\boldsymbol{B}$ (i.e., where  dense filaments are normal to the intrinsic magnetic fields). The first step to calculate the PRS (both for density and for column density distributions) is to divide the density distribution into bins (density intervals) with the same number of cells \citep[see][]{0004-637X-774-2-128}, and then $Z_{3D}$ (for density) and $Z_{2D}$ (for column density) are calculated inside each bin. Since at very high densities we have fewer structures, the last bin may include structures that have very different values of density and therefore, may have very different alignment with the field which may be hindered  by averaging. As we will see below, the HRO analysis can help to distinguish the presence of perpendicular filaments in the densest regions.

The different colors of the lines in Figure \ref{fig_3D_all_hros} indicate 3 different snapshots at which each simulation was analyzed (t=500c.u., 600c.u.,  and 700c.u.). For both sub-Alfvénic and super-Alfvénic sets, as time goes by, the density variations are only caused by compression and rarefaction due to turbulent motions. $Z_{3D}$ values do not change much along time, but they do change with different $\mathcal{M}_A$ and $\mathcal{M}_s$ values. For $\mathcal{M}_A = 0.7$, stronger magnetic fields become dynamically more important and force the motion of the turbulent flow along the magnetic field lines, thus increasing the values of $Z_{3D}$ at higher densities, specially as the sonic Mach number increases causing the formation of smaller denser structures. This does not occur at same degree in the super-Alfvénic case.

While the density distribution shows only a change in the values of $Z_{3D}$ as the Alfvénic Mach number changes (Figure \ref{fig_3D_all_hros}), for the column density along a given LOS, this behavior is realized only partially in $Z_{2D}$ (in Figures \ref{fig_2D_all_hros_godunov_Ma0.7} and \ref{fig_2D_all_hros_godunov_Ma2.0}). 

\begin{figure*}[!ht]
\centering
\includegraphics[width=1.9 \columnwidth,keepaspectratio]{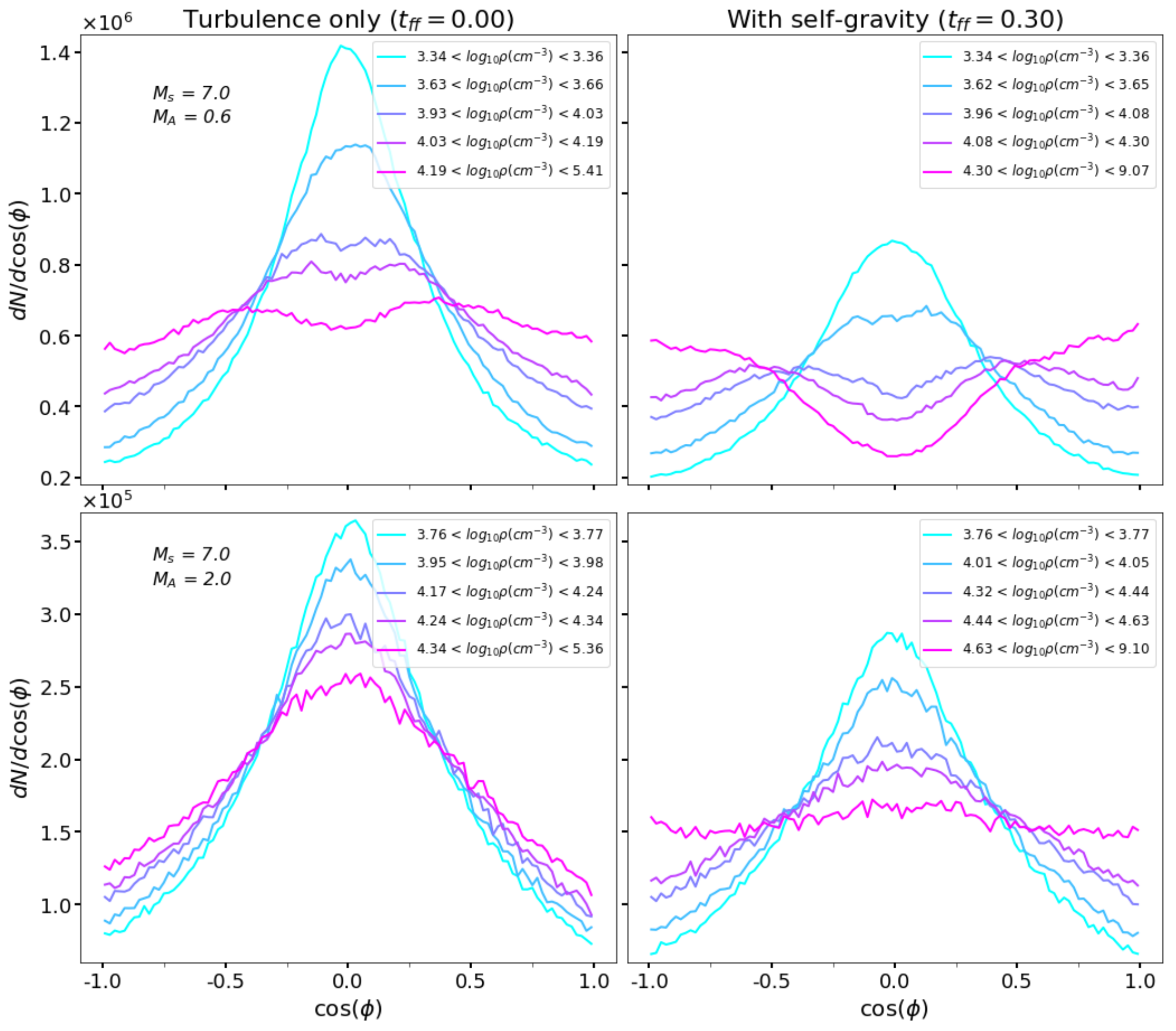}
\caption{HRO Histograms of $\cos(\phi)$ (Eq. \ref{eqtn_HRO_phi}, Section \ref{subsec:hro}) for different bins of density for the sub-Alfvénic (left) and super-Alfvénic (right) models without self-gravity. Only bins $1^{st}$, $10^{th}$, $17^{th}$,$18^{th}$,$19^{th}$ and $20^{th}$ are shown.} 
\label{fig_3D_HRO_Ms7_grav_compare}
\end{figure*}

To exemplify the distribution of angles at different density bins, Figure \ref{fig_3D_HRO_Ms7_grav_compare} shows the histograms of relative orientations (HRO); (Eq. \ref{eqtn_HRO_phi}, Section \ref{subsec:hro}) for two different models, \textit{Turb\_Ms7.0\_Ma0.7} (left) and \textit{Turb\_Ms7.0\_Ma2.0} (right). With $\phi$ being the angle between the density gradient 
and the local magnetic field (Eq. \ref{eqtn_HRO_phi}), $\cos(\phi)=0$ means that $\boldsymbol B$ is perpendicular to the density gradient while $\cos(\phi)=\pm1$ indicates that these two vectors are parallel. 

Both plots in Figure \ref{fig_3D_HRO_Ms7_grav_compare} show 6 density bins for each model. In the sub-Alfvénic case the distribution of $\cos \phi$ shows a clear peak around zero for every bin, except for the densest one, where the histogram is almost flat, meaning that there are more or less the same number of structures parallel and perpendicular to the field lines at this density bin (this is compatible with our previous analysis of Figures \ref{fig_2D_all_hros_godunov_Ma0.7} and \ref{fig:LIC_Ma0.6_turb}). At the same time, when we look to the super-Alfvénic model,  there is a higher count of $\cos \phi$ around zero, i.e. with the structures mostly aligned to $\boldsymbol B$ at all densities, even  for the densest regions (which is also compatible with the previous analysis).

The change in alignment at different densities, specially for the sub-Alfvénic models,  reflects in the PRS analysis, as there are less negative values for $Z_{3D}$ as we go to denser regions (Figure \ref{fig_3D_all_hros}). However, these regions are not numerous enough nor big enough to bring $Z_{3D}$ to positive values. 

\subsubsection{Models with self-gravity}\label{subsec:selfgrav_sims}

In models with no self-gravity the turbulent motion is the main agent modifying the magnetic field distribution. In this section we now investigate the effects of self-gravity on the orientation of the magnetic field and density.

\begin{figure}[!h]
\begin{center}
\includegraphics[width=1.05\columnwidth,angle=0]{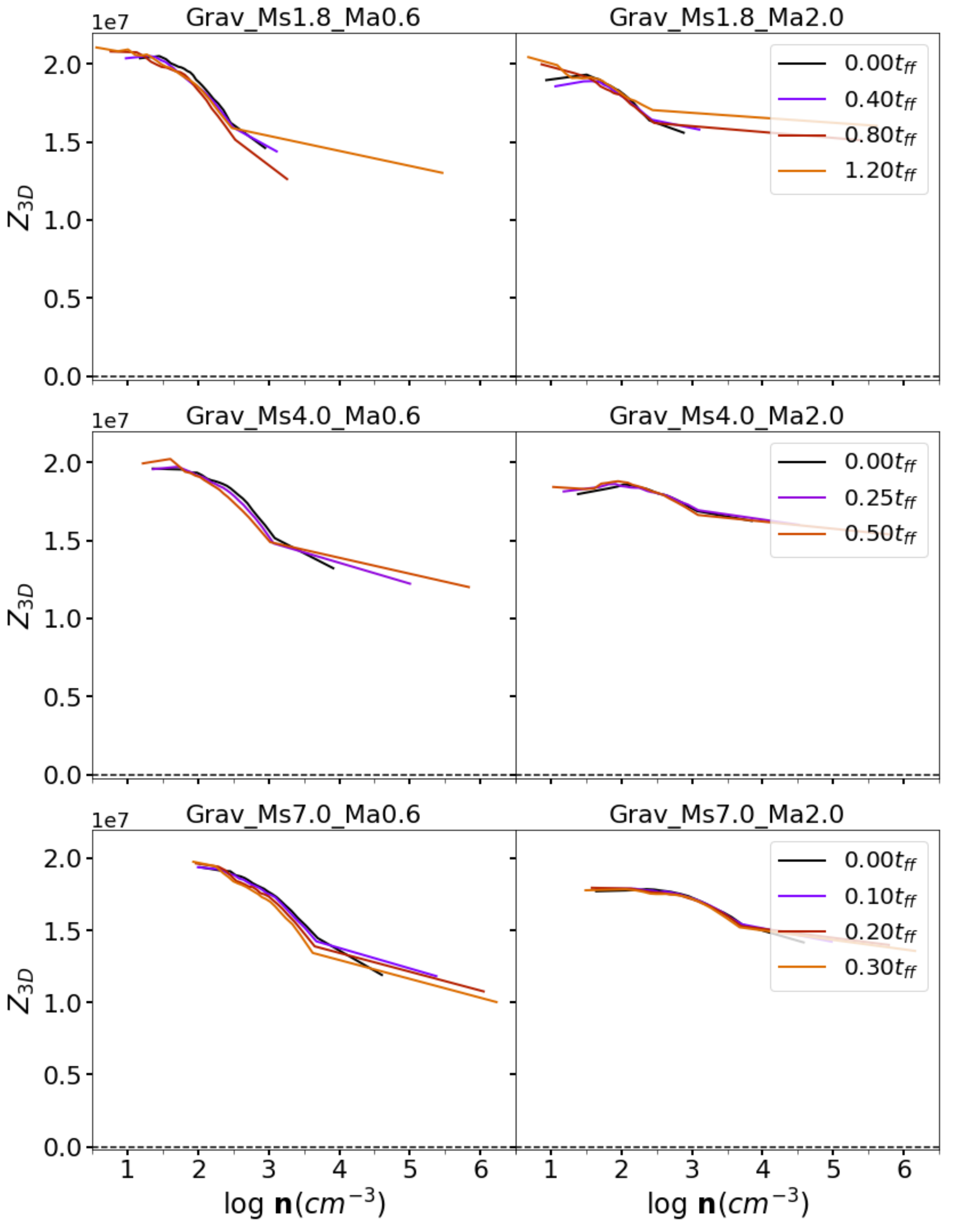}
\caption{$Z_{3D}$ analysis (Equation \ref{eqtn_PRS_3D}) applied to the 3D distribution of density and magnetic fields for all simulated models with self-gravity (see Table \ref{table_initial_param}), for different snapshots depicted in the inset. On the left we present the sub-Alfvénic (initial $\mathcal{M}_A=0.6$) and on the right, the super-Alfvénic (initial $\mathcal{M}_A=2.0$) models. From top to bottom initial $\mathcal{M}_s= 2.0, 4.0$ and $7.0$, respectively. The green dashed line indicates the critical density for star-formation (Eq. \ref{eqtn_rho_crit_PN}).}
\label{fig_ZB_grav}
\end{center}
\end{figure}

Figure \ref{fig_ZB_grav} shows the calculated values of $Z_{3D}$ for all models. As the self-gravitating regions accrete, the density gradient  becomes less perpendicular to the magnetic field at denser regions (i.e., $\boldsymbol B$ tends to become more perpendicular to these collapsing regions). This result is similar to what was seen in the previous section (Figure \ref{fig_3D_all_hros}) when no self-gravity was present. The addition is that, higher densities are achieved as time passes. As the fluid streams more easily along the magnetic field lines (since in the normal direction magnetic pressure gradients will inhibit the motion and provide support against gravity), dense structures will accumulate by gravity action mainly perpendicularly to the field direction. This can happen for both, the sub and super-Alfvénic cases, since at smaller scales magnetic fields become more and more important, as they are brought along with the collapsing regions, but is more pronounced in the sub-Alfvénic cases.

The slopes of the curves presented in Figure \ref{fig_ZB_grav} also behave in a similar way as in Figure \ref{fig_3D_all_hros}, changing for different $\mathcal{M}_s$ in the sub-Alfvénic case, while in the super-Alfvénic case, the slope does not change much. As in Figure \ref{fig_3D_all_hros}, the  values of $Z_{3D}$ are very high, but the change of slope  indicates that there is an important contribution of dense regions where $\boldsymbol{\nabla}\rho$ is predominantly parallel to $\boldsymbol{B}$, at later times. This is of course, due to the action of gravity creating collapsed regions to where the flow of matter converges. Compared to the models with no self-gravity where filaments are formed by compression forces only, in the models with self-gravity we see that lower density regions are still dominated by the interplay between turbulent motions and the magnetic field while higher density regions become dominated by an interplay between the action of gravity and magnetic fields. 

Similarly to the left panel of Figure \ref{fig_3D_HRO_Ms7_grav_compare}, the right panel shows the HRO curves for two self-gravitating models, $\textit{Grav\_Ms7.0\_Ma0.6}$ and $\textit{Grav\_Ms7.0\_Ma2.0}$. This was evaluated over the last output of these models, when $t=0.3t_{ff}$ (see the right side of Figure \ref{fig_3D_Ms7.0_compare} which shows the filamentary 3D distributions for these models). Compared to the models that do not include self-gravity, there is an enhancement in the number of regions perpendicular to the magnetic field at higher densities. With the action of gravity, even the super-Alfvénic model shows a change in the number of counts of $\cos \phi = \pm 1.0$. 

The critical density (vertical green dashed line in Figure \ref{fig_ZB_grav}) seems to be related to the densest bins, for models with sonic Mach numbers $\mathcal{M}_s= 4.0$ and $7.0$. There is a $jump$  in the values of density between the penultimate and last points in the diagrams that comes from the density range considered in the bins (as can be seen in Figure \ref{fig_3D_HRO_Ms7_grav_compare}, the $20^{th}$ bin has a wider range of densities compared to the $19^{th}$). In the super-Alfvénic case, the penultimate point is very close to $\rho_c$ for all times considered, while in the sub-Alfvénic case the penultimate point approaches $\rho_c$ as the system evolves.

Tracing the critical density of a system using the PRS is an interesting possibility, and the exact relation between the two can be further explored following the evolution of the alignment between structures and the magnetic field at smaller scales and at latter times of the collapse. However due to the lack of an adaptive mesh with increasing resolution in the densest regions in our models, this is out of the scope of this work.

Still, the results discussed up to  this point are very similar to what was shown in the previous section for models without self-gravity. We note that self-gravity does increase the number of denser regions where $\boldsymbol{B}$ is parallel to the gradient of density in the sub-Alfvénic models, but it has little or no effect when it comes to the simulations with $\mathcal{M}_A = 2.0$. At smaller scales this is probably not true, since we expect that magnetic fields should be brought along with the fluid during collapse, at some point these cores must become sub-Alfvénic and once again $\boldsymbol{B}$ would influence how the gas collapses. However, the simulations do not have enough resolution to follow the process up to this point.

\begin{figure}[!h]
\begin{center}
\includegraphics[width=1.05 \columnwidth,angle=0]{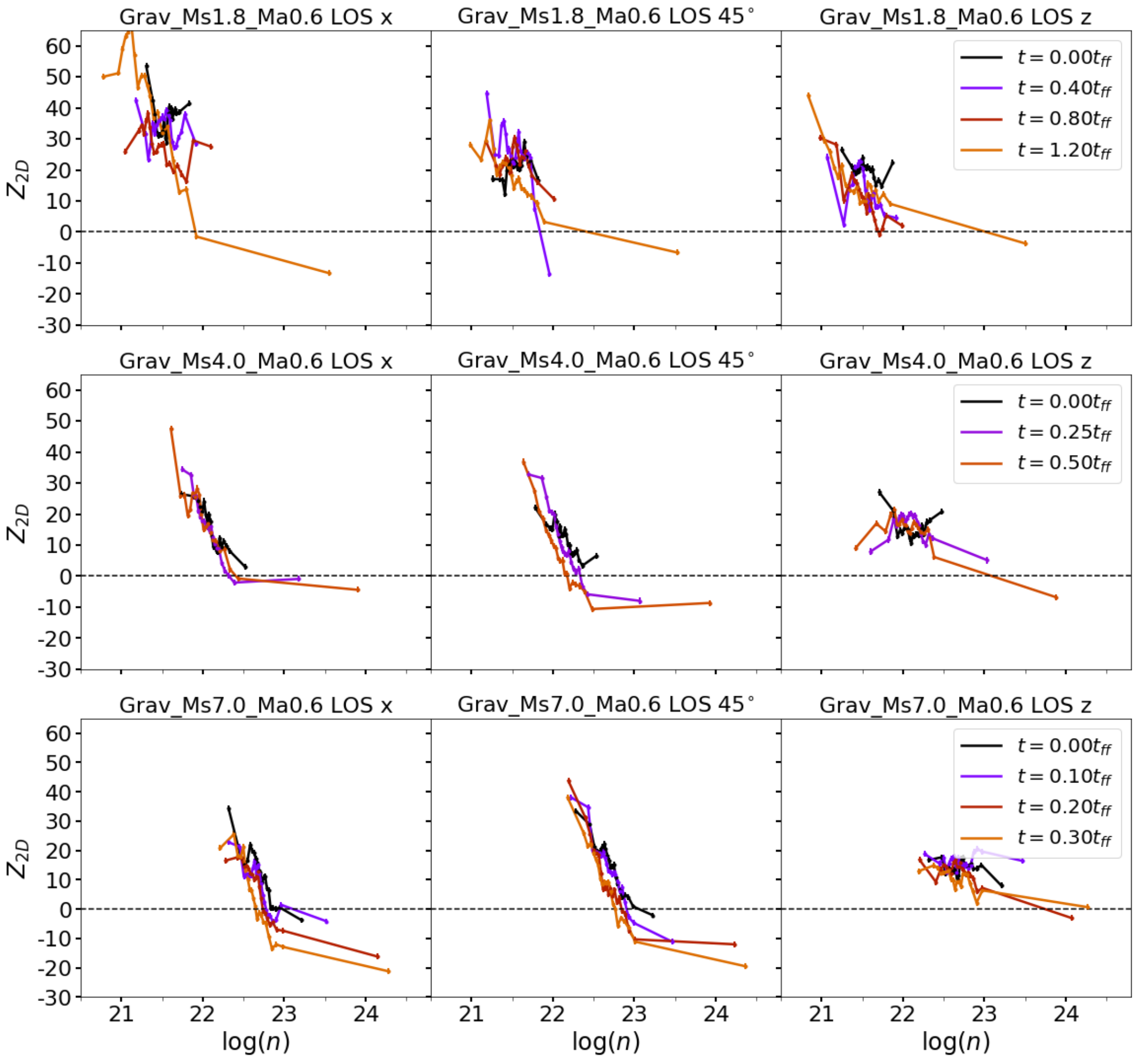}
\caption{PRS time evolution for all sub-Alfvénic (with $\mathcal{M}_A=0.6$) models with self-gravity. From left to right the PRS is applied along X (the direction perpendicular to the initial field), $45^\circ$ with regard to the initial field and Z (parallel to the initial field) directions. From top to bottom initial $\mathcal{M}_s= 2.0, 4.0$ and $7.0$, respectively. }
\label{fig_Z_2D_Ma0_6_grav}
\end{center}
\end{figure}


In order to compare the simulations with observations we need  once again to integrate along a defined LOS. We will follow same directions used in section \ref{subsec:turb_sims}.

Figures \ref{fig_Z_2D_Ma0_6_grav} and \ref{fig_Z_{2D}_Ma2.0_grav} show the PRS analysis ($Z_{2D}$) for the integrated density (column density) distribution along the three different LOS. As in Figures \ref{fig_2D_all_hros_godunov_Ma0.7}  (for initial $\mathcal{M}_A = 0.7$) and \ref{fig_2D_all_hros_godunov_Ma2.0} (for initial $\mathcal{M}_A = 2.0$), the first column represents the LOS perpendicular to the initial magnetic field, the middle column shows $Z_{2D}$ for a LOS making an angle of $45^\circ$ with respect to the initial field, and the right column, is for a LOS parallel to the initial magnetic field. From top to bottom, each line has, respectively, $\mathcal{M}_s \sim 1.8, 4.0$ and $7.0$. 

As we have seen before, the fragmentation and collapse of the structures in the cloud depends on the sonic Mach number, the higher, the faster the collapsing very dense regions appear, and the PDF of density reaches a power-law tail with a slope $\alpha = 1.5$ (see Section \ref{sec:general_models}). First, in the sub-Alfvénic cases (Figure \ref{fig_Z_2D_Ma0_6_grav}), the initial distribution of $Z_{2D}$ (black curve) is similar to the simulations without self-gravity. However, as the densest regions collapse, the effect seen in $Z_{3D}$ (Figure \ref{fig_ZB_grav}) is more pronounced for $Z_{2D}$.  

From the sub-Alfvénic models that do not consider self-gravity (see Figure \ref{fig_2D_all_hros_godunov_Ma0.7}) we saw that for higher sonic Mach numbers, and when the LOS is not parallel to initial magnetic field, $Z_{2D}$ decreases as the density increases. That is exactly what is seen in the models from Figure \ref{fig_Z_2D_Ma0_6_grav} when t = 0.0 $t_{ff}$. However, with gravity acting over the system all models evolve to a similar distribution of $Z_{2D}$, with lower densities having positive ($\boldsymbol B$ aligned with the filaments) values and higher densities showing negative values ($\boldsymbol{B}$ perpendicular to the filaments). This turns out to be the case even when the LOS is parallel to the initial field (where we see less of the original magnetic field orientation and more of the random component, see the last column of Figure \ref{fig_Z_2D_Ma0_6_grav}). In particular, for $\mathcal{M}_s = 1.8$, this trend is not initially present, but the action of gravity results in negative values of $Z_{2D}$ at later times for all LOS. In summary, self-gravity does affect the $Z_{2D}$ distribution of dense regions of sub-Alfvénic models more clearly than in models with no gravity, as we should expect, and may help to distinguish between different observed IS regions (see Section \ref{sec:comparison_obs}).

For the super-Alfvénic case ($\mathcal{M}_A = 2.0$; see Figure \ref{fig_Z_{2D}_Ma2.0_grav}) the scenario is different, the polarization vector $\boldsymbol{E}$ appears mainly aligned to the column density gradient $\boldsymbol{\nabla}N_H$ (i.e. $Z_{2D} > 0$ always, Eq. \ref{eqtn_PRS}), which means that  the projected magnetic field is more frequently parallel to the structures. This is similar to the  super-Alfvénic models without self-gravity  (Figure \ref{fig_2D_all_hros_godunov_Ma2.0}; see also Figure \ref{fig:LIC_Ma2.0_evolution}), but in Figure \ref{fig_Z_{2D}_Ma2.0_grav} larger column densities and $Z_{2D}$ values are achieved.

\begin{figure}[!h]
\begin{center}
\includegraphics[width=1.05 \columnwidth,angle=0]{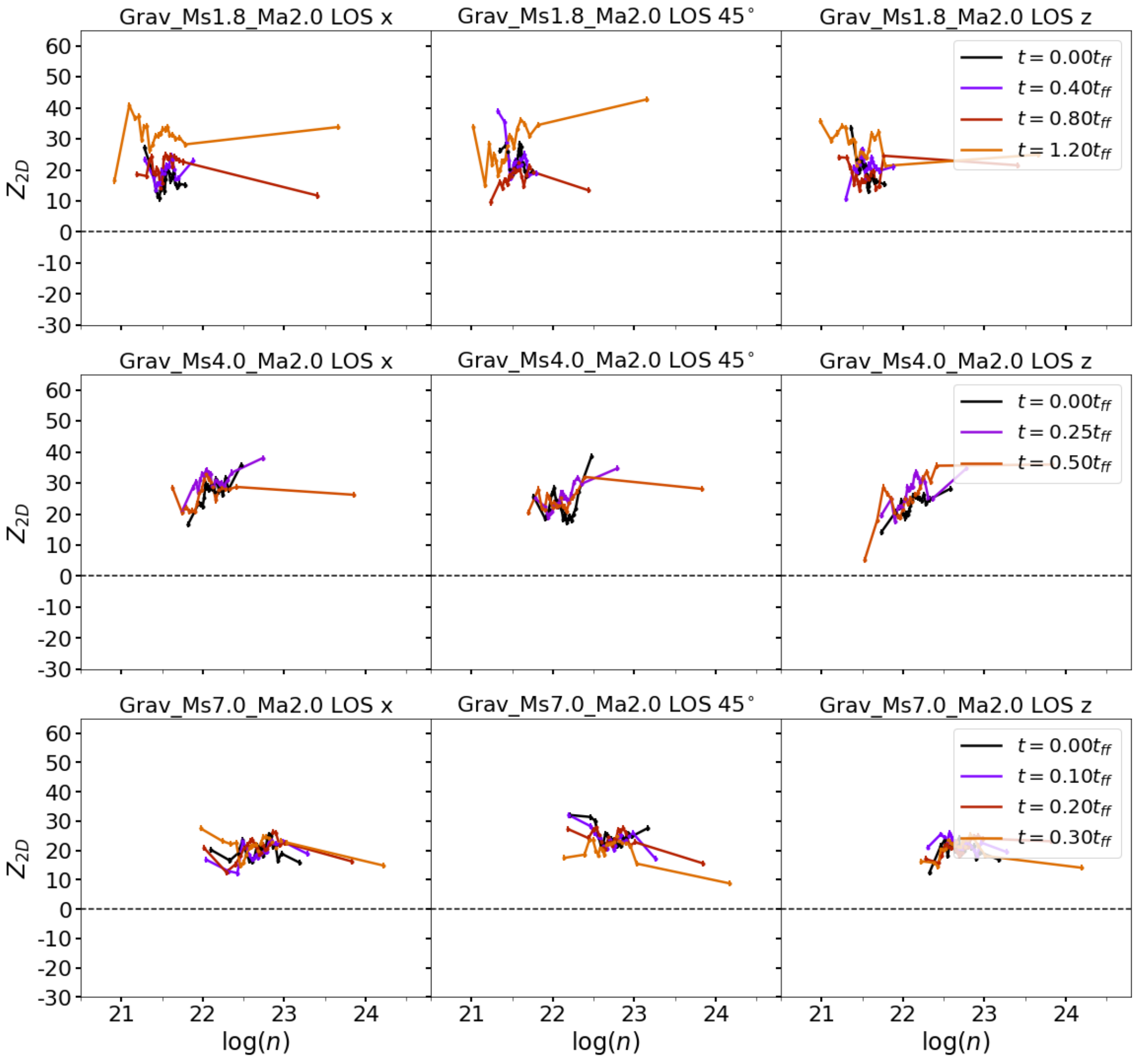}
\caption{PRS time evolution for all super-Alfvénic (with $\mathcal{M}_A=2.0$) models with self-gravity. From left to right the PRS is applied along X (the direction perpendicular to the initial field), $45^\circ$ with regard to the initial field and Z (parallel to the initial field) directions. From top to bottom initial $\mathcal{M}_s= 2.0, 4.0$ and $7.0$, respectively.}
\label{fig_Z_{2D}_Ma2.0_grav}
\end{center}
\end{figure}

\section{Comparison with observations}\label{sec:comparison_obs}

To compare the diagnosis of the simulated models presented in the previous Section with observations we use clouds observed by \textit{Planck Satellite} \citep{2016AA...586A138P} and BLASTPol \citep{2017A&A...603A..64S}. These objects have been also analyzed by \citet{2018MNRAS.474.1018J}.

The \textit{Planck Satellite} observed thermal emission and dust polarization in 7 bands between 30 and 353 GHz. In the case of \citet{2016AA...586A138P}, the observations were made using the High Frequency Instrument at 353 GHz. The molecular clouds reported have distances estimated between $\sim 150$ pc and $\sim 450$ pc. Given their angular sizes of about $15^\circ$, these distances imply that the clouds have sizes up to $\sim100$ pc. The sonic Mach number estimated for these regions may vary. For instance,   Polaris MC has $\mathcal{M}_s$ varying between $\sim3-7$ depending on the region inside the cloud. Orion has $\mathcal{M}_s \sim 8$ \citep{2013ApJ...766L..17S}. \citet{2017AA...608L...3K} also reports comparable sonic Mach measured from CO lines for Ophiucus, Taurus and Musca.  Temperatures range between $\sim10-30$K \citep{2013MNRAS.432.1424K}. Our models are compatible with these values.

Vela C observations were obtained using BLAST-Pol to estimate the magnetic field direction \citep{10.1117/12.2054759,2016ApJ...824..134F,2016ApJ...824...84G}, and \textit{Heschel}, to derive the column density maps \citep{2011A&A...533A..94H}. BLAST-Pol used three wavelengths centered at 250, 350 and 500 $\mu m$. For the column density, \textit{Heschel}, SPIRE and PACS data were used, with observations made at 160 (PACS), 250, 350 and 500 $\mu m$. Previous studies place the cloud around a 700 pc distance, with a total mass of more than $10^5$ \(\textup{M}_\odot\). Despite being a massive cloud, it is still at an early stage of evolution, and some authors claim that only one or two O-type stars have been formed \citep{2017A&A...603A..64S}. The estimated temperature ranges between $\sim10-30$K inside the cloud \citep{2011A&A...533A..94H}.

\subsection{PRS and angular distribution analysis}

\begin{figure*}[!ht]
\begin{center}
\includegraphics[width=1.5 \columnwidth,angle=0]{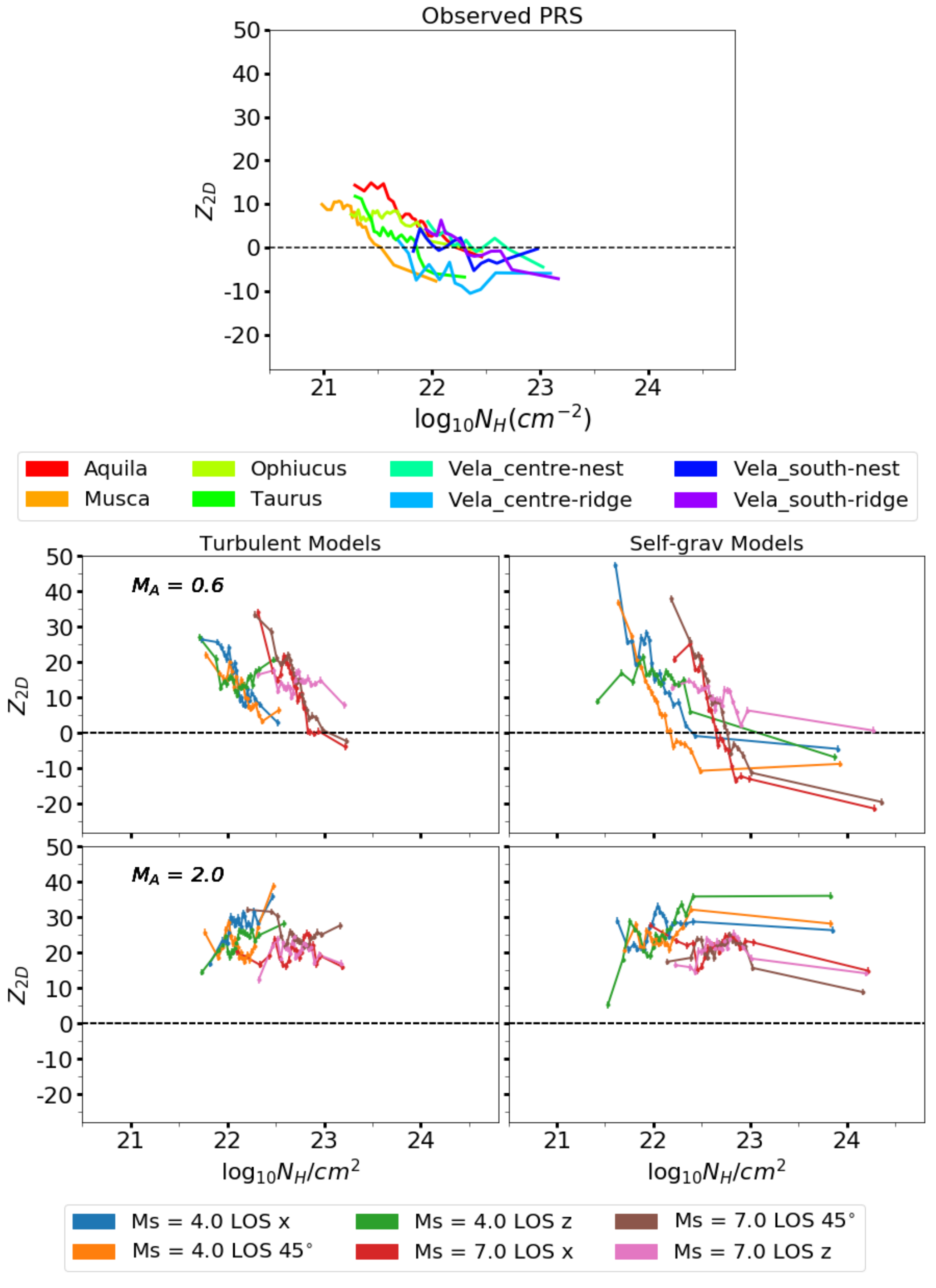}
\caption{Top: $Z_{2D}$ (Eq. \ref{eqtn_PRS}) calculated from observations \citep{2016AA...586A138P,2017A&A...603A..64S}. Bottom: $Z_{2D}$ (Eq. \ref{eqtn_PRS}) calculated from our simulated models considering all LOS for models with $\mathcal{M}_s= 4$ and $7$. The PRS when only turbulence is present is  on the left. The PRS calculated for the final snapshot of models that consider self-gravity is shown on the right. The Alfvénic Mach number of the models is indicated above each plot.} 
\label{fig_obs_compare_PRS}
\end{center}
\end{figure*}

Figure \ref{fig_obs_compare_PRS} shows the comparison of the PRS presented in Section \ref{subsec:selfgrav_sims} with the results obtained for Vela C and a few of the clouds observed by \citet{2016AA...586A138P}. 

The PRS analysis is sensitive to the number of points in the sample. This means that directly comparing the values of $Z_{2D}$ with the ones obtained in observations requires some caution and we discuss this later in this section. Nevertheless, our results offer some insight into the magnetic field and what kind of structure distribution we can expect inside the observed MCs.

First, we discuss the general distribution of the projected magnetic field on the sky ($\textbf{B}_{\perp}$) for most clouds. In general,  $\textbf{B}_{\perp}$ appears parallel to a single direction inside these molecular clods, with a smaller fraction being randomly distributed. Chamaleon-Musca and Aquila, for instance, are good examples of this behavior \citep[see Figure 3, left, from][]{2016AA...586A138P}. As discussed in the previous section, this is a characteristic observed mainly in our sub-Alfvénic models.

From the integrated maps obtained from our supersonic, sub-Alfvénic models (Figures \ref{fig:LIC_Ma0.6_turb} and \ref{fig:LIC_Ma0.6_grav}), $B_\perp$ becomes more chaotic as the LOS gets closer to the direction of the mean magnetic field. The LOS then catches the effect of regions where locally the field can appear as a random twist, while still showing a general coherence for most of the cloud. The presence of such a characteristic in the distribution of the observed $\boldsymbol B_\perp$ is and indication that these clouds are sub-Alfvénic.

Evidence in favor of a sub-Alfvénic description of the turbulence in these clouds is the PRS analysis presented in Figure \ref{fig_obs_compare_PRS}. \citet{2018MNRAS.474.1018J} have revised the data presented in the works mentioned above in order to apply the PRS method described in Section \ref{subsec:hro}. This reveals that the observed relation between density gradients and the projected magnetic field in the sky is similar to the sub-Alfvénic models we simulated, going from positive to negative as density grows, i.e. less dense regions appear more aligned to the magnetic field, while dense regions appear more perpendicular to it\footnote{Note that this trend appears consistently only in sub-Alfvénic models.}. For most clouds, the turbulent models approximately produce a behavior of $Z_{2D}$ compatible with the observations. The different regions of Vela C, in particular, show higher column density values and most of the structures appear perpendicular to the projected magnetic field, since they have a negative tail of $Z_{2D}$ at higher densities. This is only achieved in the sub-Alfvénic models that consider self-gravity (see Figure \ref{fig_obs_compare_PRS}). Aquila, on the other hand, can be described by a sub-Alfvénic model without self-gravity. This conclusion is supported by the PRS data and the information from the observed column density. Aquila does not show large gradients of column density, which is characteristic of our turbulent models without self-gravity. Moreover, the cloud is known for having star formation activity inside isolated pockets \citep{2008hsf1.book...18P,2008hsf2.book..693E}, and since the PRS data was calculated using a bigger portion of the cloud, it makes sense that the observed PRS follows a distribution similar to our turbulent models.


Complementary information is provided in Figures 3 and 4 from \citet{2016AA...586A138P}, that present the observed integrated column densities of the clouds Aquila, Chamaeleon-Musca, Taurus and Ophiucus on the left side, and the histograms of relative orientations between the projected magnetic fields and the density gradient of the structures in these clouds. Comparing with the HROs of our models in Figure \ref{fig_3D_HRO_Ms7_grav_compare}, we note a similar behavior with the sub-Alfvénic models, i.e., as we go from less dense to denser structures, the relative orientation between the magnetic field and the filaments goes from aligned to perpendicular, particularly in the case of Chamaeleon-Musca and, at some extent, Taurus. Note that Figure \ref{fig_3D_HRO_Ms7_grav_compare} shows the HRO applied to the 3D structures and not to column density maps, still the behavior is similar.

One effect that can also influence the comparison of the PRS from observations with our simulated models is related to the resolution and field of view that is possible to achieve with the telescopes. It is important to highlight that due to self-gravity, the units considered in the code are scale dependent. The size of the clouds in our simulations is 10 pc, while the clouds observed can extend up to hundreds of parsecs. This means that our results are more representative of sub-structures inside the clouds and not so much of the global formation of the cloud and their surroundings. The PRS results reported in \citet{2018MNRAS.474.1018J} are for entire regions observed by the \textit{Planck satellite}. Even though, our study presents results that are qualitatively comparable to the observations and hence, one can always argue that due to the self-similarity nature of the turbulent clouds, the general behavior at very large scales does not differ much of that in the intermediate scales inside the clouds, at least in scales where self-gravity is not dominant yet \citep{bialy2020}.  
To account for the scaling effect above, we may take the  self-gravity, sub-Alfvénic simulations with $\mathcal{M}_s = 1.8$. 
The Poisson equation considers the normalized potential:

\begin{equation}\label{eqtn_poisson_eq_norm}
    \nabla^2 \Psi' = \nabla^2 \Big( \frac{\Psi}{4 \pi G} \Big) = \rho.
\end{equation}

The normalization implies a gravitational constant in code units as follows:

\begin{equation}\label{eqtn_Gconst_norm}
    G_{c.u.} = G_N \rho_0 \left( \frac{L}{c_s} \right)^2,
\end{equation}

\noindent where $G_N$ is the gravitational constant, $\rho_0$ is the initial density of the simulation, and $L$ the size of the domain. With this in mind, it is possible to re-scale our models as long as we keep $G_{c.u.}$ the same, i.e. the ratio in the right-hand side of the equation, $\rho_0 \left(L/c_s \right)^2$, must be kept constant (see alternative ways of scaling gravity keeping constant the virial parameter in e.g., Mckee et al. 2010).

\begin{figure*}[!ht]
\begin{center}
\includegraphics[width=1.95 \columnwidth,angle=0]{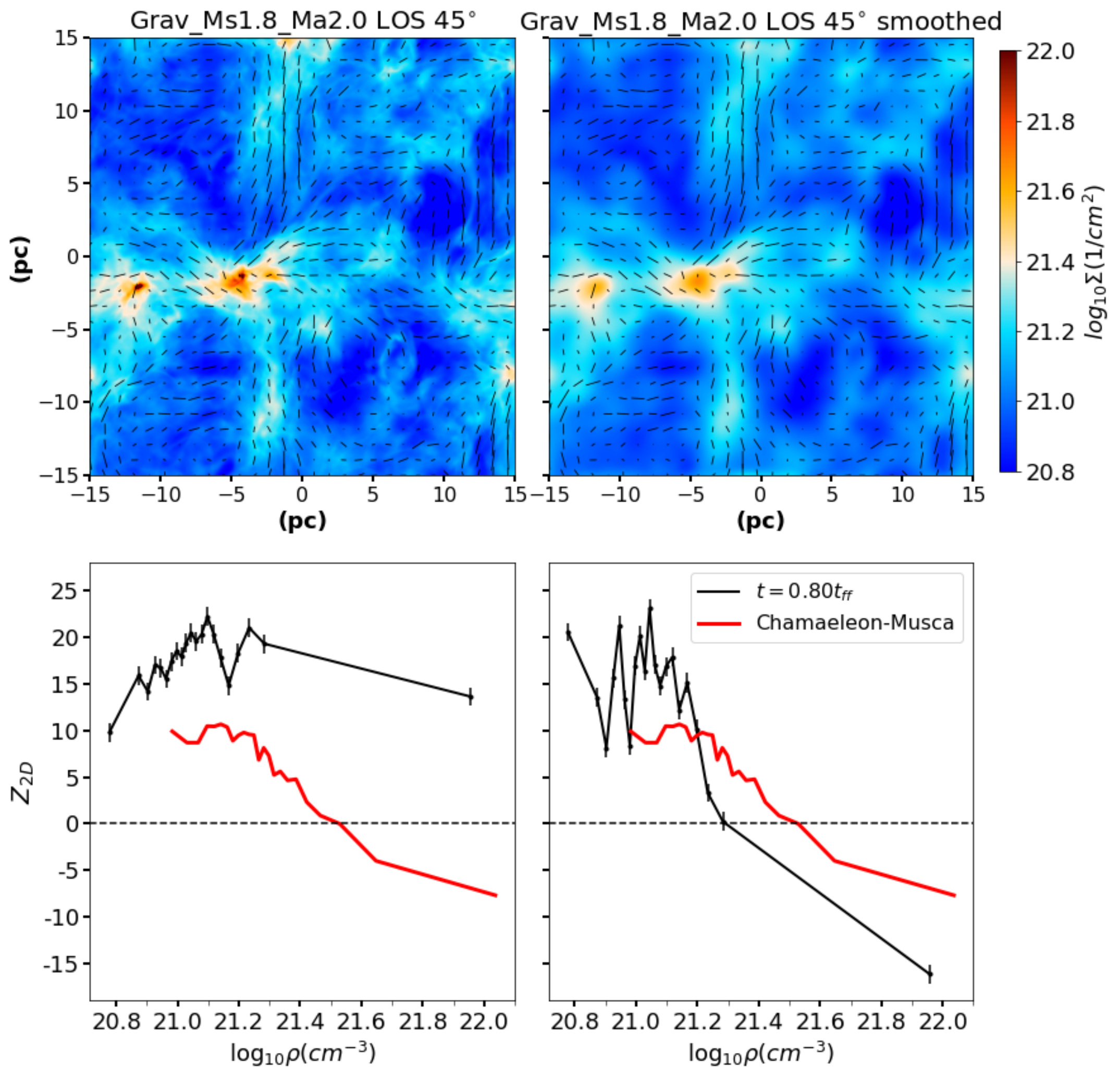}
\caption{Left: Column density map along a LOS $45^\circ$ inclined with respect to the initial magnetic field for our model \textit{Ms1.8\_Ma2.0\_grav} (top) and the PRS calculated for the respective map (bottom). Right: Same map, but convolved with a Gaussian kernel in a similar processes to the one made for observations from \citet{2016AA...586A138P} (top) and the PRS of the respective map (bottom);\citep[see also][for further details]{0004-637X-774-2-128}.} 
\label{fig_Ms1.8_smoothed}
\end{center}
\end{figure*}

We consider a region of 40 pc, which is approximately the estimated extension of the  clouds observed by \citet{2016AA...586A138P}. Considering the same temperature for the re-scaling ($10$K), the average density in this larger system is around $6\,cm^{-3}$ to keep the ratio $\rho_0 \left(L/c_s \right)^2$ constant. The final result is shown in Figure \ref{fig_Ms1.8_smoothed}.

On the left side of Figure \ref{fig_Ms1.8_smoothed}, the original column density and PRS that were extracted from the simulation are shown for the re-scaled system. The integration was made along a direction $45^\circ$ inclined with respect to the initial magnetic field. On the bottom left diagram of Figure \ref{fig_Ms1.8_smoothed}, the red line is the obtained PRS  from observations of Chamaeleon-Musca (same as in the top diagram of Figure \ref{fig_obs_compare_PRS}), and the black line is the PRS calculated for model \textit{Ms1.8\_Ma\_2.0\_grav}. The right-hand side of the figure shows the same column density map of the left, but convolved with a 2D Gaussian kernel \citep[see][for further details]{0004-637X-774-2-128}. The image has been smoothed  roughly to the same spatial resolution of the observations, as it was done by \citet{2016AA...586A138P}. In our simulation, the smoothing length corresponds to about 7 cells. We note that the PRS for the smoothed map is quite different at higher densities, even if the column density is not much different from the original one, thus providing different information. In fact, we see that the PRS calculated for the smoothed map is more similar to the observed one (red curve). Intrinsically (i.e., examining the high resolution map of the left), these structures are aligned to the magnetic field for all column densities (explaining the positive PRS), but with the loss of resolution at the smaller scales of the observations (in the map of the right), the densest  structures actually appear perpendicular to the magnetic field. Both maps  share similarities with Chamaeleon-Musca, but the alignment indicated by the PRS in the smoothed map, while being compatible with the observations, is actually not representative of the real behavior of the projected magnetic field onto the sky (as suggested by the simulation of the left panels).

\subsection{A closer look inside the Molecular Clouds}

Observations made by \citet{2013A&A...550A..38P} using \textit{Herschel} Telescope have revealed several smaller structures around the filament B211/3 in Taurus molecular cloud. The dense filament (B211/3) appears perpendicular to the magnetic field around it, while less dense structures \citep[the striations observed by][]{2013A&A...550A..38P} are parallel to the projected magnetic field onto the sky. According to the scale indicated in the Figure, the size of the region is about $\sim3\,pc \times 4\,pc$. The separate regions of Vela C as defined in \citet{2017A&A...603A..64S} also have similar sizes.

\begin{figure*}[ht]
\begin{center}
\includegraphics[width=1.2 \columnwidth]{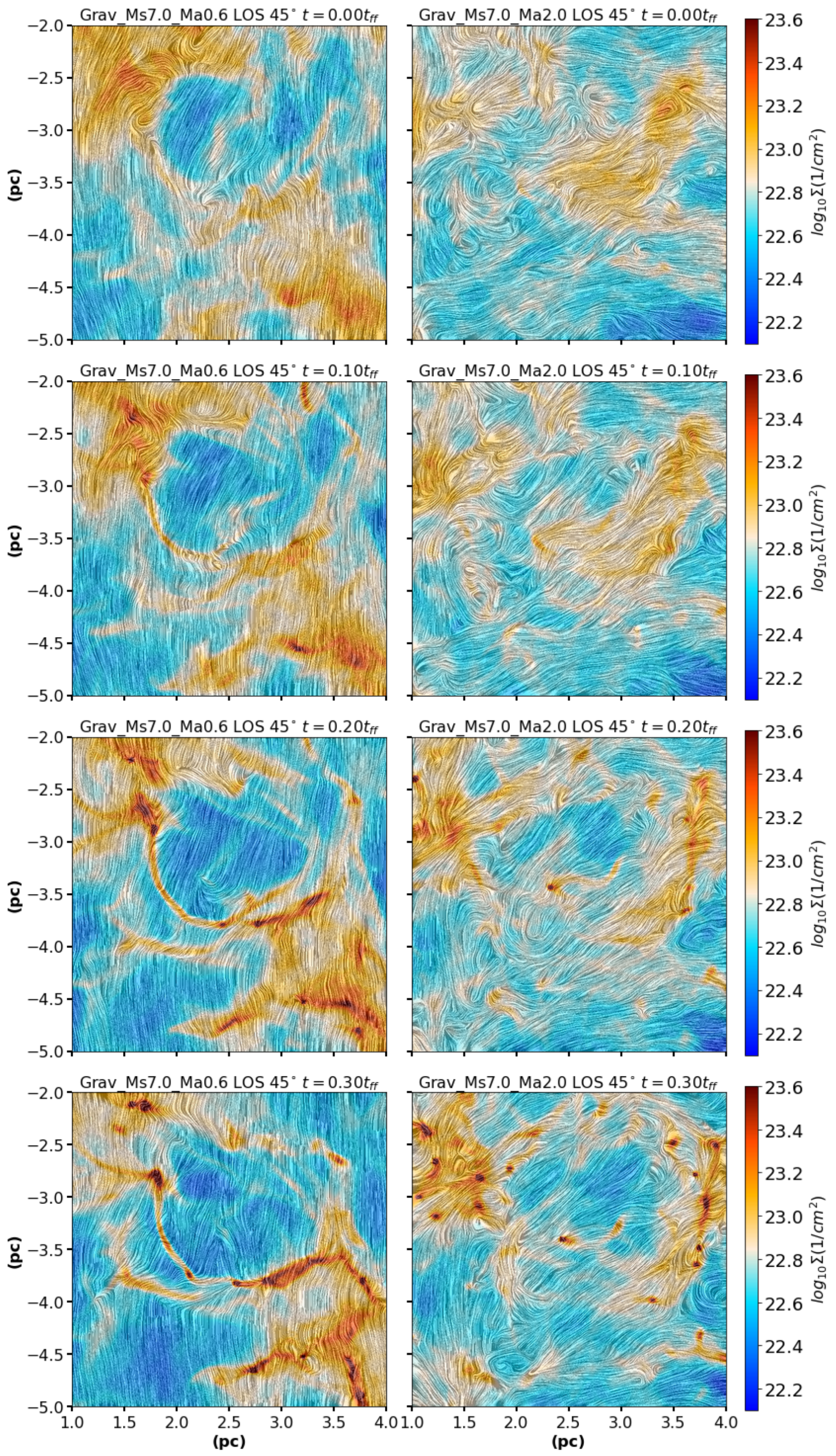}
\caption{Time evolution of zoomed-in regions extracted from models \textit{Ms7.0\_Ma0.6\_grav} (left) and \textit{Ms7.0\_Ma2.0\_grav} (right). Both maps were integrated along a LOS $45^\circ$ inclined with respect to the initial magnetic field. The time considered is indicated above each map.} 
\label{fig_Ms7.0_slice}
\end{center}
\end{figure*}

To evaluate the behavior of these smaller, denser regions, both of Taurus and Vela C, Figure \ref{fig_Ms7.0_slice} shows the time evolution of the column density of two regions of similar size to these clouds extracted from our self-gravity models with $\mathcal{M}_s = 7.0$. The density integration was along a LOS making an angle of $45^\circ$ with respect to the original magnetic field. On the left, we have $\mathcal{M}_A = 0.6$, on the right $\mathcal{M}_A = 2.0$ and the time is indicated on the top of each image.  While in the sub-Alfvénic case filaments appear mostly perpendicular to the projected magnetic fields onto the sky, in the super-Alfvénic case they commonly appear aligned with the magnetic fields.

\begin{figure*}[ht]
\begin{center}
\includegraphics[width=1.7 \columnwidth,angle=0]{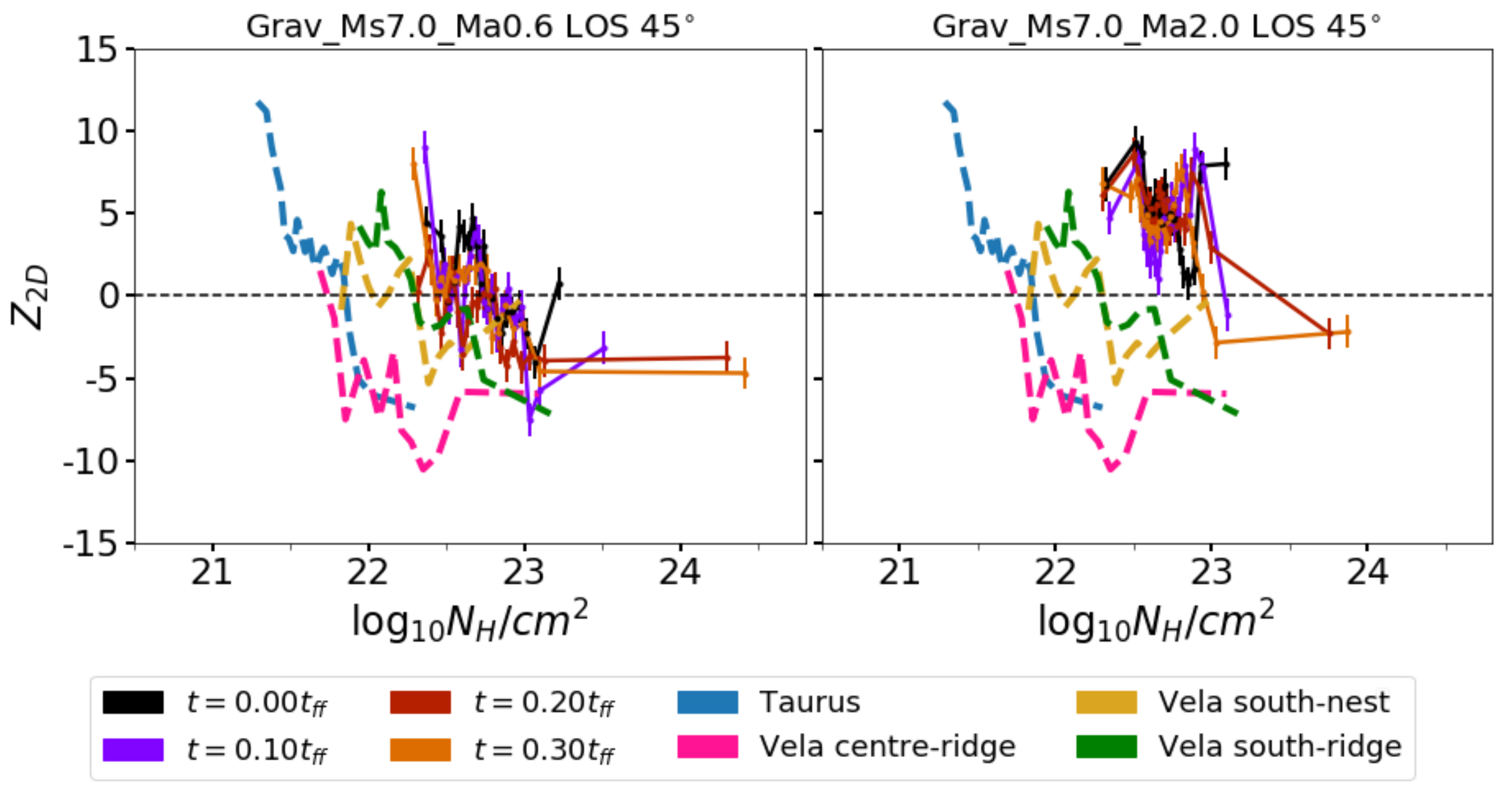}
\caption{Time evolution of the PRS analysis ($Z_{2D}$, Eq. \ref{eqtn_PRS}) for the same maps presented in Figure \ref{fig_Ms7.0_slice}. Solid lines represent the PRS along time calculated from our models, sub-Alfvénic on the left-hand side and super-Alfvénic on the right hand side. Dashed lines represent the observed PRS of some regions presented in the top diagram of Figure \ref{fig_obs_compare_PRS}.} 
\label{fig_Ms7.0_Zx2D_slice}
\end{center}
\end{figure*}

In Figure \ref{fig_Ms7.0_slice}, initially (at $t=0.0t_{ff}$), it is possible to see filaments both parallel and perpendicular to the magnetic field in the sub-Alfvénic model. At this time, as only turbulence and magnetic fields are present, the compression motions tend to align the filaments with the magnetic lines (an effect which is more pronounced in the super-Alfvénic model on the right side of the figure), while the stronger magnetic fields imposed by the sub-Alfvénic regime tend to oppose resistance to alignment, through their tension and pressure gradient forces. As time passes, matter flows along the lines and results in denser filaments perpendicular to the field. In the super-Alfvénic case, the magnetic field is dragged with the flow resulting in projected magnetic fields aligned to the filaments. Along time, it is also possible to see dense regions where the field is perpendicular to the structures, which is evidenced in Figure \ref{fig_Ms7.0_Zx2D_slice} for $t=0.3t_{ff}$, with $Z_{2D}<0$ for the highest column densities in both models.

In Figure \ref{fig_Ms7.0_Zx2D_slice}, the solid lines show the PRS analysis of the column density maps presented in Figure \ref{fig_Ms7.0_slice}. The results for the sub-Alfvénic models are shown on the left-hand side, and the ones for the sub-Alfvénic model are shown on the right-hand side. The dashed lines show the PRS calculated for some of the observations (see top diagram of Figure \ref{fig_obs_compare_PRS}) as a comparison. 

Once again the distribution of $Z_{2D}$ obtained in our sub-Alfvénic models seems a better representation of what is observed in the sky. Note that the values of $Z_{2D}$ for the super-Alfvénic case when $t=0.3t_{ff}$ and several regions have already collapsed, also show negative values for higher column densities, which is expected as discussed in Section \ref{sec:prs_hro_results}, but the distribution for smaller densities and at previous times is not compatible with the observation. On the other hand, the sub-Alfvénic model shows very similar behavior when compared to the observations. In particular, the PRS analysis obtained from the sub-Alfvénic model in Figure \ref{fig_obs_compare_PRS} resembles  what is seen in Vela C. 

Our models were made considering only solenoidal driving turbulence. However as discussed in Section \ref{sec:intro}, compressive modes may change the distribution of the filaments and this may affect the resulting PRS analysis. We will study this in a forthcoming work. 

\section{Discussion}\label{sec:dicussion}
In this section we  compare our results with previous studies and  summarize our  findings. 
\subsection{Comparison of our results with previous studies in the literature}

\citet{0004-637X-774-2-128} have first analyzed the alignment of structures with the magnetic fields in a molecular cloud using statistical tools like those employed in this work. However, in their study turbulence was not constantly driven in the simulated system, and thus was allowed to decay with time. Also, they did not consider sub-Alfvénic models and investigated only a single sonic Mach number. Their highest magnetized model had $\mathcal{M}_A=3.16$ and $\mathcal{M}_s=10$, so that they could not investigate most relevant dynamical effects of the magnetic fields in the evolution of star forming systems, as in the present work. Compared to our most similar model, \textit{Ms7.0\_Ma2.0\_grav}, their results present significant differences. They only consider a LOS perpendicular to the initial magnetic field and find a distribution of $Z_{2D}$ that is closer to our distributions for sub-Alfvénic models. Also, the general coherence of $\boldsymbol{B}_\perp$ in their column density maps is only observed in our sub-Alfvénic models. This difference is most likely due to the fact that turbulence was not continuously driven. As turbulence decays their system  approaches the sub-Alfvénic regime, with gravity and magnetic pressure becoming the main forces acting on the fluid. Hence, the collapse primarily occurs along the field lines, resulting in dense structures perpendicular to the magnetic field. As turbulence is dominated by the magnetic field it is unable to bend the lines, thus explaining why their results are more comparable with our sub-Alfvénic models.

In a more recent study, using the same set of simulations as in the work above, \citet{2017A&A...607A...2S} derived an expression for the time evolution of the angle  between the density gradient and the magnetic field in the turbulent MCs and found that  these two quantities are preferentially  either  parallel or perpendicular to each other. In our simulations, we identify a similar trend only for the evolved sub-Alfvénic models with self-gravity (see Figure \ref{fig_Z_2D_Ma0_6_grav}). In addition, these authors have  concluded that  the observed change in the relative orientation between column density structures and the projected magnetic field in the plane of sky, from mostly parallel at low column densities to mostly perpendicular at the highest column densities, would be the result of  gravitational collapse and/or convergence of flows. This trend is also identified in our models, specially in the sub-Alfvénic ones with self-gravity (see Figures \ref{fig_Z_2D_Ma0_6_grav} and \ref{fig_Ms7.0_Zx2D_slice}). 

Our results are also in accordance to those reported by  \cite{2020arXiv200300017S}. As in our models without gravity, they do not observe perpendicular structures to the mean field, particularly in their lower magnetization models. As we have seen, a sub-Alfvénic cloud observed in a LOS parallel to the magnetic field and a super-Alfvénic cloud under the action of gravity can yield similar results with regard to the general alignment between magnetic field and filaments.

\citet{2017ApJ...842L...9H} have also performed 3D MHD simulations in order to study the alignment at smaller scales in order to compare with observations made by ALMA. As initial conditions, they considered a single sonic Mach number ($\mathcal{M}_s=10$) and different Alfvénic Mach number cases, including a trans-Alfvénic and a sub-Alfvénic one. The general setup of their models is similar to the ones presented in this paper, with column density maps also comparable to ours, but their analysis focus on the formation of collapsed cores. They have studied the morphology of the magnetic field around these cores for different scenarios in order to compare to ALMA observations. In our work however, we have focused  on the formation of filaments and the interaction of these structures with the magnetic fields. Both works can thus be seen as complementary to each other, with the results presented here giving some insight about the birthplace of cores that will ultimately form stars. 

\citet{2018MNRAS.480.2939G} studied the structure of magnetic fields inside self-gravitating filaments in turbulent environments. They note that the magnetic field around the filament is primarily perpendicular to the structure and the collapse along the filament would later bend the magnetic field lines creating U-shapes. However, they argue that the lack of resolution, as well as the decrease of polarization in observations would not allow to detect these features. This is consist  with the result we found in Figure \ref{fig_Ms1.8_smoothed}, where the decrease of resolution, simulated by the smoothing of the image, revealed a complete change of the behavior  in the PRS.

A final remark regarding the  result from Figure \ref{fig_Ms1.8_smoothed} is in order. Though  applied to Chamaeleon-Musca cloud, it has served to illustrate how  re-scaling may influence the results obtained from observations. In particular, the similarities seen in the general behavior of $\boldsymbol{B}_\perp$ and the formation of larger collapsed regions perpendicular to it, as well as the comparison of the PRS showing a reasonable match in Figure \ref{fig_Ms1.8_smoothed}, indicate that this issue should be further explored in future works.

\section{Conclusions}\label{sec:conclusion}

In this work, we  have  explored, by means of 3D MHD simulations of isothermal MCs, how the relative alignment between  density structures  and  the magnetic fields is affected by different regimes of MHD turbulence and  by self-gravity. We have also examined how projection effects of these structures on the plane of sky may be compared with polarization observations of distinct clouds.

We considered models with initially homogeneous magnetic field and density, and different sonic Mach numbers ($\mathcal{M}_s\sim2.0,4.0$ and $7.0$) and Alfvénic Mach numbers ($\mathcal{M}_A\sim0.7$, and $\mathcal{M}_A\sim2.0$), in a $10\,pc \times 10\,pc \times 10\,pc$ volume. To ensure the gravitational collapse of our clouds in the presence of self-gravity, we considered an initial  ratio between turbulent and gravitational energy density  $\alpha_{vir} \sim 0.5$ (Eq. \ref{eqtn_alpha_vir}).

Our evolved turbulent models with no self-gravity show a PDF distribution compatible with a lognormal (Eq. \ref{eqtn_lognorm_pdf}), and a density power-spectrum compatible with  supersonic motions. While the models with self-gravity, as structures collapse,  develop a power-law distribution at higher densities which becomes shallower with time in agreement with previous studies. Their density power-spectrum  also deviates from the initial distribution becoming flatter with time. This means that the changes in the statistics, e.g. the PRS, at smaller and dense regions of these models are due to the action of gravity, which ultimately helps creating perpendicular structures to the magnetic field and decrease or increase the value of $Z_{2D}$ or $Z_{3D}$, respectively.  

Computing the histograms of the relative orientations between the magnetic fields and the density gradients and applying the PRS analysis \citep{2018MNRAS.474.1018J} enables better understanding of the interaction of density structures with the magnetic field inside molecular clouds. Considering the results from the analysis of the column density maps with models that do not consider self-gravity (Section \ref{subsec:turb_sims}) we have found that:
    \begin{itemize}
        \item The LOS in the the sub-Alfvénic model cases changes the PRS distribution ($Z_{2D}$; Equation \ref{eqtn_PRS}), and this is also affected by the magnitude of the sonic Mach number. For smaller sonic Mach numbers ($\mathcal{M}_s\sim2.0$) most of the polarization vector $\boldsymbol{E}$ appears parallel to the column density gradient  ($\boldsymbol \nabla N_H$), $Z_{2D} >0$ in all models, that is, the magnetic fields are mainly aligned to the filaments, due to action of supersonic turbulent compression motions. For larger sonic numbers, the fragmentation increases and smaller values of $Z_{2D}$ can be seen for denser regions (specially if the LOS is not parallel to the initial magnetic field). This indicates that these denser smaller regions have a greater contribution of structures perpendicular to $\boldsymbol B_\perp$, due to the dominance of the magnetic forces that prevent gas motion across the lines, facilitating their flow along them.
        
        \item For super-Alfvénic models, the less intense magnetic field does not show the same coherence along large length scales, since the lines are more easily twisted by the turbulent motions, and the column density maps show similar characteristics at different LOS. The PRS analysis also reflects this, with $Z_{2D}$ presenting only positive values (often ranging between $\sim5-20$) for all LOS. Different sonic Mach numbers also show no clear effect in the alignment of the structures with the field.
    \end{itemize}{}

When gravity is included these scenarios change.

For the models with self-gravity (Section \ref{subsec:selfgrav_sims}) we have found that:
    \begin{itemize}
        \item 
        There is an enhancement of dense structures perpendicular to the projected magnetic field to the sky, even when looking along the mean field. When the system is sub-Alfvénic, smaller values of $Z_{2D}$ are present for all LOS at higher column density values. At later times, all models show at some degree a change from positive to negative values in $Z_{2D}$ which implies the presence of dense, collapsing structures with magnetic fields normal to them, as one should expect, since the collapse is easier along the magnetic fields. 
        
        \item For the super-Alfvénic models, $Z_{2D}$ is positive for most LOS, indicating that most structures are aligned with the projected magnetic field to the sky. This effect is less prominent for $\mathcal{M}_s = 7.0$, where more fragmentation again propitiates the formation of denser collapsed cores inside filaments, and since magnetic fields are aligned to them, these cores appear as perpendicular structures to the field and therefore yield lower values of $Z_{2D}$.
    \end{itemize}{}

We conclude that self-gravity can create structures perpendicular to the magnetic field, even for mean super-Alfvénic Mach numbers. As evidenced also by the PDF and power spectrum, dense regions of MCs are clearly dominated by gravity, while less dense regions are mainly affected by turbulence, as lower densities still sit in the lognormal branch of the PDF (see Figure \ref{fig_PDFs_Ms7.0_compare}). Effects of projection due to the LOS may change the observed alignment for less dense regions (see Figure \ref{fig_Z_{2D}_Ma2.0_grav}), but still overdense regions show smaller values of $Z_{2D}$ (this effect can be seen by comparing the bottom panel of Figure \ref{fig_Z_{2D}_Ma2.0_grav} with the right panel of Figure \ref{fig_Ms7.0_Zx2D_slice}).

The observed behavior of the PRS in column density maps is a result of the intrinsic distribution of filaments and magnetic fields inside the molecular clouds. With regard to the 3D analysis of the density structures, we have found that:
    \begin{itemize}
        \item For the sub-Alfvénic models, the inclusion of gravity helps the creation of structures perpendicular to the strong magnetic field, but supersonic turbulence favors the formation of less dense filaments parallel to the field (see Figure \ref{fig_3D_Ms7.0_compare}). This is more easily realized through the histograms of $\cos(\phi)$ (top diagrams of Figure \ref{fig_3D_HRO_Ms7_grav_compare}), with $\cos(\phi) = \pm 1$ having a higher number of counts for the bin of largest density. In accordance to what is observed in the column density maps for $Z_{2D}$ (Figures \ref{fig:LIC_Ma0.6_turb} and \ref{fig:LIC_Ma0.6_grav}), the PRS analysis of the density shows smaller values of $Z_{3D}$ with higher sonic Mach numbers, specially at higher densities.
        
        \item For the super-Alfvénic models, magnetic field lines appear mostly aligned to filaments when only turbulence is considered, with the sonic Mach having little or no effect. But models with gravity have the densest structures  perpendicular to the magnetic field  at later times.   
        
    \end{itemize}{}

Finally, in Section \ref{sec:comparison_obs} we have compared the results described above with observations made by \textit{Planck}, \textit{Herschel} and BLASTPol. The comparison indicates that, qualitatively, our sub-Alfvénic models can better reproduce the characteristics of observed clouds. Not only the behavior of the observed $Z_{2D}$ , but also the general coherence of the magnetic field projected on the plane of the sky ($\boldsymbol B_\perp$), is compatible with our sub-Alfvénic models for most clouds. There are clouds where twists of $\boldsymbol{B}_\perp$ could be explained with effects due to LOS. Clouds like Aquila, for instance, can be well represented by models with no self-gravity or in earlier stages of collapse, while Taurus and Vela C have some similarities with the models with a more advanced stage of gravitational collapse.

\acknowledgments
L.B-M. acknowledges support from the Brazilian funding agency CAPES and also the kind hospitality and  partial support  of the Simons Foundation Flatiron Institute during  his visit in July 2018; E.M.G.D.P. acknowledges support from the Brazilian funding agencies FAPESP (grant 2013/10559-5) and CNPq (grant 308643/2017-8);  B.B.  is grateful for the support of the Simons Foundation Flatiron Institute and the Packard Foundation.  RSL is supported by FAPESP (2013/10559-5) and LHSK also acknowledges support from  FAPESP (2016/12320-8).  We are grateful for the use of MHD turbulent boxes from the Catalog for Astrophysical Simulations (CATS; www.mhdturbulene.com).
Part of the numerical simulations presented here were performed in the cluster of the Group of Plasmas and High-Energy Astrophysics (GAPAE), acquired with support from FAPESP (grant 2013/10559-5). This work also made use of the computing facilities of the Laboratory of Astroinformatics (IAG/USP, NAT/Unicsul), whose purchase was also made possible by FAPESP (grant 2009/54006-4), and of the Simons Foundation Flatiron Institute. 

\section*{Data Availability}
The data of this article will be available upon request to the corresponding author.

\newpage
\bibliography{hro-mhd.bib}

\appendix

\section{\\Appendix A: Resolution effects}\label{sec:apendixA}

\begin{figure}[!h]
\begin{center}
\includegraphics[width=0.98 \columnwidth,angle=0]{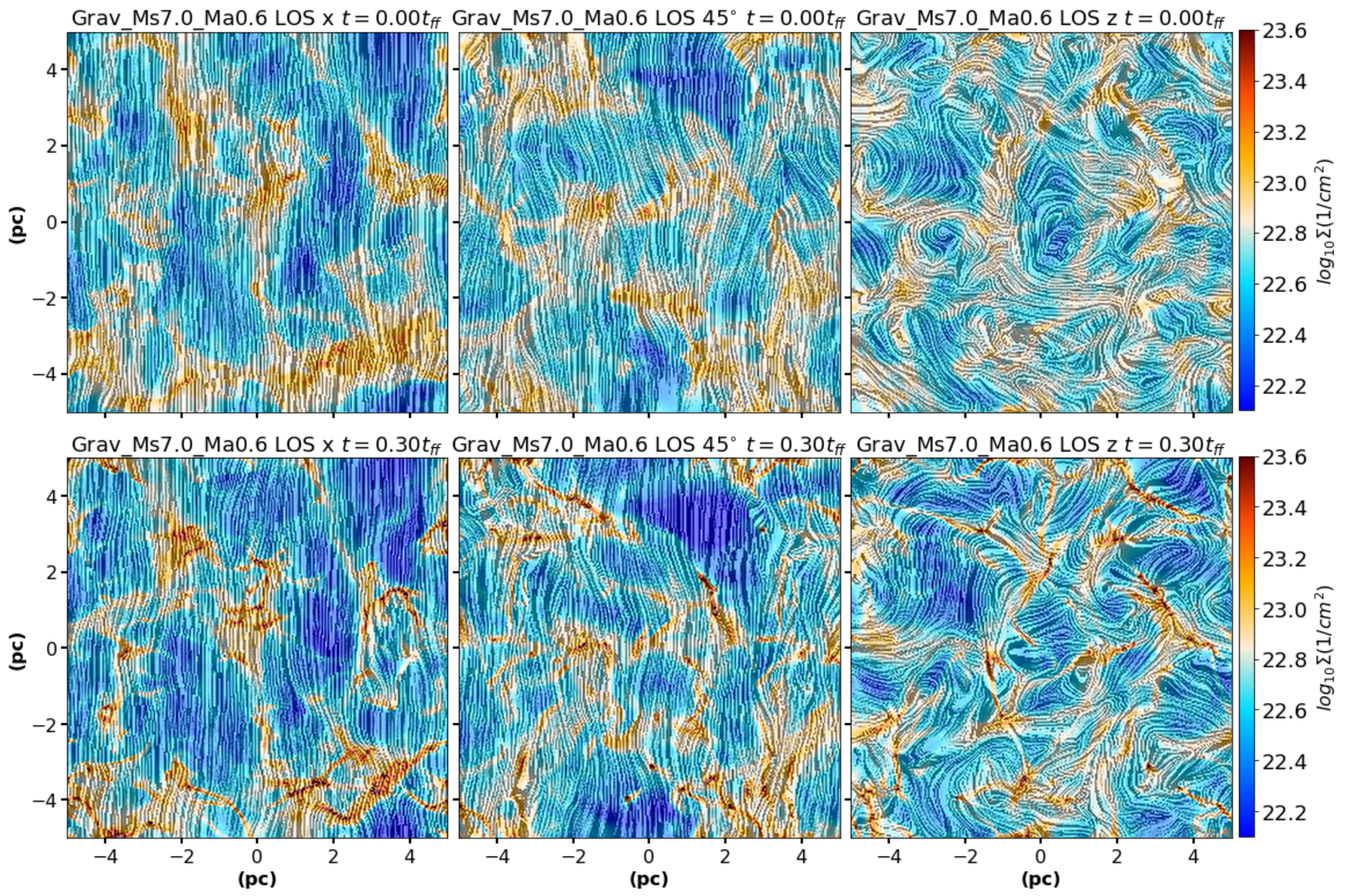}
\caption{Column density maps with LIC method applied to $\boldsymbol{B}_\perp$ for  sub-Alfvénic models with $\mathcal{M}_A= 0.6$ and sonic Mach number $\mathcal{M}_s = 7.0$ with resolution $256^3$, for three different LOS.  Top row shows the column density map at the initial snapshot  ($t=0.0t_{ff}$) (with fully developed  turbulence  and no self-gravity), for each LOS. Bottom panel  shows the final snapshot (when self-gravity becomes important) for the same LOS. From left to right, the column density distribution is integrated along X (the direction perpendicular to the initial field), $45^\circ$ with regard to the initial field, and Z (parallel to the initial field) directions.}
\label{fig:LICmap_256}
\end{center}
\end{figure}

\begin{figure}[!h]
\begin{center}
\includegraphics[width=0.98 \columnwidth,angle=0]{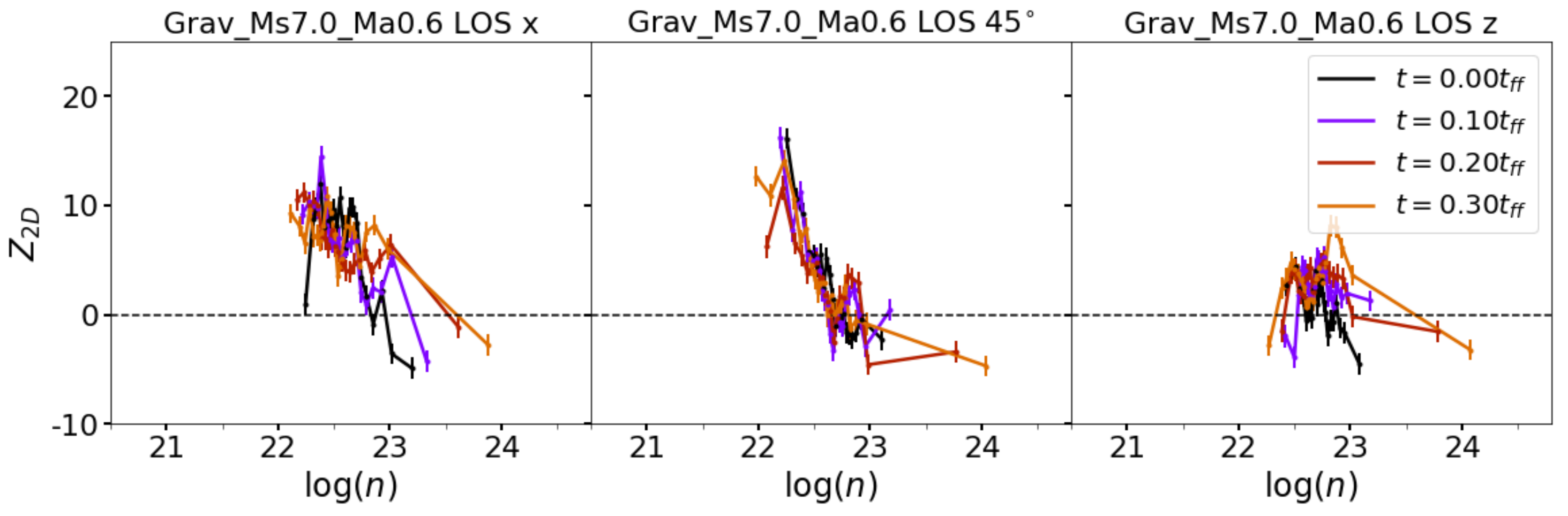}
\caption{PRS time evolution for the same sub-Alfvénic model presented in Figure \ref{fig:LICmap_256}. From left to right the PRS is applied to the LOS along X (the direction perpendicular to the initial field), $45^\circ$ with regard to the initial field and Z (parallel to the initial field) directions.}
\label{fig:Z_2D_256}
\end{center}
\end{figure}

Figures \ref{fig:LICmap_256} and \ref{fig:Z_2D_256} depict  column density maps and PRS time evolution, respectively, for the same sub-Alfvénic models with $\mathcal{M}_A= 0.6$ and  $\mathcal{M}_s = 7.0$ presented  in Sections \ref{sec:general_models} and \ref{subsec:hro}, except that here they have resolution of $256^3$. A quick comparison with the higher resolution counterparts (in Figures \ref{fig:LIC_Ma0.6_turb} and \ref{fig:LIC_Ma0.6_grav}),   shows that the column density maps are quite similar. The PRS  in Figure \ref{fig:Z_2D_256} shows some differences  relative to the higher resolution model in Figure \ref{fig_Z_2D_Ma0_6_grav}, particularly at later times when gravity has  caused the collapse of several structures and the differences due to resolution become more obvious at these smaller scales, but the general behavior discussed in Sections \ref{sec:general_models} and \ref{subsec:hro} is already present at these smaller resolution models.

\end{document}